\documentclass[,nonblindrev]{informs3} 

\OneAndAHalfSpacedXI

\usepackage{natbib}
 \bibpunct[, ]{(}{)}{,}{a}{}{,}%
 \def\bibfont{\small}%
\TheoremsNumberedThrough     
\ECRepeatTheorems

\EquationsNumberedThrough    

\MANUSCRIPTNO{}

\usepackage[flushleft]{threeparttable} 
\usepackage{booktabs}
\usepackage{array}
\usepackage{caption}  
\usepackage{float}
\usepackage{multirow}
\usepackage{CJKutf8}  
\usepackage{makecell}
\usepackage{comment}
\usepackage{subfigure}
\usepackage{amssymb}
\usepackage{bbm}
\usepackage{dsfont}
\usepackage{hyperref}
\hypersetup{ 
    colorlinks=true,
    linkcolor=red,
    urlcolor=blue,
    citecolor=blue
}
\usepackage{xcolor}

\begin{document}




\TITLE{\large Unintended Consequences of Recommender System Interventions: Evidence from a Field Experiment}
\RUNAUTHOR{Luo, Yao, and Zhang}
\RUNTITLE{Unintended Consequences of Recommender System Interventions}
\ARTICLEAUTHORS{%
\AUTHOR{Shilei Luo}
\AFF{Olin Business School, Washington University in St. Louis, \EMAIL{\url{l.shilei@wustl.edu}}} 
\AUTHOR{Song Yao}
\AFF{Olin Business School, Washington University in St. Louis, \EMAIL{\url{songyao@wustl.edu}}}
\AUTHOR{Dennis J. Zhang}
\AFF{Olin Business School, Washington University in St. Louis, \EMAIL{\url{denniszhang@wustl.edu}}}
} 

\ABSTRACT{%
Platform content interventions in recommendation systems are typically evaluated as static ``nudges,'' ignoring that the systems adaptively learn from the resulting user behavior. We investigate this dynamic through a large-scale field experiment on a short-video platform. The experiment involves a ``sleep reminder'' campaign designed to reduce late-night usage. Paradoxically, the intervention \textit{increased} late-night engagement by 14.75\% and overall platform usage by 2.18\%, and the effects persisted for weeks even after the experiment. We explain this through a forced-exploration mechanism, showing that by revealing high latent demand for the promoted content, the intervention triggers a recommendation policy update that routine user behavior would not produce. The data generated by the intervention induced the algorithm to update its post-campaign policy, reinforcing the very engagement loops the campaign aimed to mitigate. Our findings demonstrate that user-facing interventions can effectively retrain the underlying algorithm, triggering durable, system-wide shifts in content distribution that challenge standard evaluation metrics in platform governance and social responsibility initiatives.
}%



\maketitle


\section{Introduction}

Modern social media platforms deliver content primarily through sophisticated recommender systems that curate personalized feeds for billions of users \citep{ansari2002ecustomization, yoganarasimhan2020search, holtz2020engagement}. These systems have become indispensable to the platforms for strategic content promotion. For instance, platforms routinely boost influencer partnerships, amplify brand-sponsored videos\footnote{ \href{{https://www.forbes.com/sites/emilybaker-white/2023/01/20/tiktoks-secret-heating-button-can-make-anyone-go-viral/?sh=3671f4e16bfd}}{Baker-White, E. (2023). TikTok’s Secret `Heating' Button Can Make Anyone Go Viral. \textit{Forbes}.} Accessed on November 24, 2025.}, and elevate political messages by adjusting ranking scores to increase visibility \citep{huszar2022algorithmic, saveski2022}. Such interventions are now a standard practice in digital marketing, with platforms offering sophisticated targeting tools that allow content promoters to reach specific demographic segments and behavioral cohorts. However, existing research has primarily evaluated these interventions through narrow, short-term metrics tied to immediate campaign objectives, while their broader, long-term consequences on user behavior and recommender system-level algorithmic adaptation remain poorly understood. This gap exists because recommender systems are not static distribution channels; instead, they are adaptive learners that continuously update their internal representations of user preferences based on observed user behavior \citep{lambrecht2019algorithmic, jiang2019degenerate}. This creates a feedback loop where interventions shape user behavior, which in turn reshapes the algorithm's future recommendations. These interactions generate system-level dynamics that traditional evaluations, focusing narrowly on immediate responses, routinely overlook.

We study this feedback loop using a large-scale randomized field experiment on a major short-video platform. The experiment centers on a temporary platform-run content intervention. Because assignment was randomized, we can compare treated and control users during the campaign and after the intervention ended. This design allows us to measure not only immediate user behavior, but also whether recommendations and engagement changed after the platform removed the intervention. The key empirical challenge is that the intervention changes the data available to the recommender system. When users respond to content that would otherwise have been shown less often, their behavior may become an input into subsequent algorithmic updates. We use the experiment to trace the sequence from intervention-induced exposure to user response and then to post-campaign recommendation and engagement outcomes.

Our empirical setting is a celebrity-driven ``sleep reminder'' campaign, a corporate social responsibility initiative designed to reduce unhealthy late-night usage through brief videos featuring celebrities encouraging users to rest. The campaign provides a useful test because it explicitly intended to \textit{decrease} engagement. A persistent increase in engagement therefore serves as a sharp signal of the underlying algorithmic dynamics. By temporarily boosting the ranking scores of celebrity-recorded sleep reminder videos between 1 and 5 a.m., the campaign exposed treated users to content that the platform's baseline policy would have otherwise filtered out. The results were striking and counterintuitive: rather than reducing consumption, the intervention increased treated users' late-night usage by 14.75\% and overall engagement by 2.18\% during the campaign week, with positive effects on engagement persisting for multiple weeks after the campaign ended. We show that this paradoxical outcome is consistent with a preference-discovery mechanism. The intervention exposed treated users to celebrity-related content that the baseline recommender system had under-supplied for some users. Their responses to the boosted content generated new behavioral signals, and subsequent recommendations placed greater weight on related content even after the campaign had ended.

To fixate the empirical analysis, Section~\ref{sec:model} summarizes a simple conceptual framework and Appendix A presents the formal model. The model isolates three elements. First, before the campaign, routine feedback can leave the recommender system inside an \emph{inaction zone} when the system under-supplies a content category and therefore observes too little informative user behavior to justify a policy update. Second, during the campaign, the intervention changes user exposure to the content category and generates new user responses. Third, after the campaign, the boost has ended, but the recommender system may update its policy using the user responses generated during the intervention period. This structure clarifies why a temporary intervention can have persistent effects and yields empirical implications for exposure, engagement, feed reallocation, conversion patterns, and treatment-effect heterogeneity.

We organize the empirical analyses according to the conceptual framework. First, we verify that the campaign changed content exposure during the intervention period by increasing treated users' encounters with the specific content. Second, we estimate short-run and post-campaign treatment effects on late-night and overall usage, documenting that the increase in engagement continued after the boost in the content ended. Third, we use granular platform logs to examine changes in content-category exposure, conversion-rate patterns, downstream feed reallocation, and heterogeneous effects across users with different baseline exposure to the promoted content. These tests keep the randomized experiment at the center of the analysis while linking the evidence to the three-period logic of the framework: a pre-campaign incumbent policy, a campaign-period exposure shock that generates new user responses, and a post-campaign recommendation update based on those responses.

Our research makes several contributions. First, we present a large-scale randomized field experiment that traces the long-term, system-level consequences of a platform intervention. We reveal that a content moderation effort aimed at promoting social good can inadvertently backfire by making the platform more engaging. This finding contributes to the empirical literature on digital nudges and platform governance by showing that adaptive recommendation systems can transform a temporary user-facing campaign into persistent changes in engagement. Second, we identify algorithmic learning from intervention-induced behavior as a mechanism through which these unintended consequences arise. The evidence links the campaign to increased exposure, short-run and post-campaign engagement, category-level feed reallocation, conversion-rate patterns, and user-level heterogeneity, which together help distinguish recommender adaptation from pure persuasion or novelty effects. Third, we offer concrete managerial and policy implications. User-facing interventions can function as algorithmic training signals, so platforms need to evaluate them using long-term, system-level metrics rather than only immediate campaign outcomes. The same logic also suggests that carefully designed interventions can be used as strategic exploration tools to identify latent preferences and improve content-user matching.

The remainder of the paper proceeds as follows. Section~\ref{sec:litreview} reviews literature on recommender systems, algorithmic adaptation, and platform interventions. Section~\ref{sec:model} summarizes the conceptual framework, and Appendix A presents the formal model. Section~\ref{sec:setting} describes the experimental design and data. Section~\ref{sec:results} presents empirical results. Section~\ref{sec:mechanism} discusses mechanisms and interpretation. Section~\ref{sec:sutva} examines spillovers and threats to SUTVA. Section~\ref{sec:conclusion} concludes with implications for research and practice.

\section{Literature Review}\label{sec:litreview}

Our research builds upon and contributes to three distinct yet interconnected streams of literature: (1) the dynamics of recommender systems, specifically interventions and feedback loops; (2) large-scale field experiments on online platforms; and (3) marketing research on social responsibility and digital well-being.

\subsection{Recommender Systems: Interventions and Feedback Loops}

Recommender systems (RS) are fundamental in shaping user behavior on digital platforms \citep{adomavicius2005toward}. Extensive research has demonstrated how RS influences consumption and engagement. Personalization, for example, has been shown to substantially increase click-through rates, purchase propensity, and overall time spent on platforms \citep{ansari2002ecustomization, yoganarasimhan2020search, li2022recommender, holtz2020engagement}. These improvements generate billions of dollars in business values \citep{gomez2016netflix, zhou2010impact} by reducing search costs and improving the quality of matches between users and items \citep{yuan2024recommendation, ghose2014examining, chen_yao2017}. Achieving effective personalization, however, requires the system to continuously explore and learn user preferences \citep{xu2023scalable}. 

A central debate concerns the impact of personalization on consumption diversity. One perspective warns of ``filter bubbles'' \citep{berman2020curation}, arguing that RS algorithms narrow user interests and favor ``blockbuster'' products \citep{chaney2018algorithmic, liu2023daily, fleder2009blockbuster}. Conversely, others suggest RS can broaden horizons by surfacing niche ``long-tail'' content \citep{brynjolfsson2006longtail, hosanagar2014will}. Analytical work on strategic recommendation design shows that the optimal information design may involve ``overselling'' or ``demarketing'' products depending on the alignment between seller and consumer incentives \citep{berman2022strategic}. Recent large-scale field experiments suggest a complex reality: outcomes depend heavily on specific RS algorithmic designs \citep{holtz2020engagement, lee2019how}, and short-term engagement gains often diverge from long-term user welfare \citep{donnelly2023welfare}.

Our work sits at the intersection of two critical sub-streams of this literature: interventions on recommender systems and adaptive feedback loops in these systems. 

First, platforms routinely intervene in their recommender systems to achieve strategic goals, such as boosting influencer partnerships, elevating political messages, or arranging search results to shape which products users see \citep{huszar2022algorithmic,saveski2022,lam2021platform,bairathi2025value}. Even targeted advertising can function as an implicit recommendation that shapes consumer beliefs, sometimes leading firms to paradoxically choose less precise targeting \citep{ning2025targeted}. However, existing research has primarily evaluated these interventions through narrow, short-term metrics, often overlooking the system's adaptive dynamic response.

Second, recommender systems are not static; they continuously learn and evolve through feedback loops where RS algorithm's outputs shape user behavior, which in turn feeds back into the model as training data for subsequent model updates. Existing literature often highlights the externalities of these feedback loops, including embedding systematic biases \citep{lambrecht2019algorithmic,iyer2024competitive}, perpetuating unfairness in the labor market \citep{cowgill2019bias, fu2022un}, amplifying political polarization \citep{huszar2022algorithmic}, distorting market prices \citep{fu2022does}, and triggering algorithm aversion when recommendations conflict with users' own experience or peer behavior \citep{liu2023aversion}. When optimized solely for short-term engagement, RS algorithms risk becoming trapped in suboptimal equilibria.

Our study bridges these two streams by conceptualizing the ``sleep reminder'' intervention as a form of ``forced exploration''. This perspective connects to recent work on how the explore-exploit tradeoff operates in platform environments and batch campaign settings, where information externalities across users within the same campaign fundamentally alter the value of exploration \citep{li2023seasonal,rashid2025auctions}. Unlike previous studies that focus on how feedback loops propagate errors and biases \citep{sandvig2014auditing}, we demonstrate a correction process. Namely, the intervention compels the system to ingest fresh high-signal data, allowing the system to correct ``embedding bias,'' meaning a misalignment between the system’s prior beliefs about users' preferences and the users' true preferences. This process enables the algorithm to escape suboptimal equilibria and improve long-run content-user matching.

\subsection{Field Experiments on Online Platforms}

Our empirical methodology is grounded in the literature on large-scale field experiments in marketing and economics. A/B testing has become the ``gold standard'' for evaluating platform interventions \citep{gordon2019comparison, kohavi2020trustworthy, blake2015field}. Researchers use experiments to compare algorithmic and editorial news curation \citep{peukert2023editor}, study the effects of new platform features \citep{moehring2022news}, analyze algorithmic pricing's effect on inequality \citep{zhang2021can}, assess the impact of platform design on economic outcomes \citep{einav2016peer}, improve experimental sensitivity \citep{deng2013improving}, and advance the evaluation of targeting policies \citep{simester2020efficiently}.

A recurring and critical theme in this research stream is the prevalence of unintended consequences. Seminal work, such as the ``Music Lab'' experiment \citep{salganik2006experimental}, demonstrates how minor social cues can cascade into arbitrary market outcomes. More recent evidence similarly documents that intended-direction campaigns can produce reverse behavioral outcomes in digital environments \citep{liaukonyte2023spilling}. This risk is particularly relevant in the context of ``digital nudges'' \citep{thaler2008nudge}. While designed to steer behavior benevolently, these interventions can backfire if they fail to account for the complexity of user interactions, network effects, peer spillovers \citep{aral2012identifying, eckles2016estimating}, or the endogenous system-level responses they trigger \citep{godes2009firmwom}. More broadly, interventions designed to improve market or social outcomes can generate countervailing responses from downstream participants that partly reverse the policy goal \citep{kim2022government}. Recent platform experiments also show that relatively small platform-level interventions can generate both immediate and enduring changes in user engagement \citep{lu2023immediate}.

We contribute to this literature by identifying a new source of complexity: the \textit{endogenous adaptive nature of the algorithm} itself. While prior studies have emphasized how \textit{users} react to interventions by the system, we highlight how the \textit{system} reacts to the interventions. The large-scale randomized field experiment demonstrates that standard A/B testing frameworks, which typically measure direct, short-term user responses, are insufficient for capturing the dynamics due to the reactions of both users and the system. By tracking outcomes over two months, we show that the algorithm’s learning process can generate long-run, system-wide consequences that dominate and even reverse the intervention’s intended effects. Consequently, our work offers a methodological template for future research, advocating for experimental designs with longer time horizons that capture the dynamic co-evolution of human behavior and machine learning.

\subsection{Social Responsibility in Marketing}

Finally, our research speaks to the growing literature on corporate social responsibility (CSR) and platform governance. A central theme of this stream explores how ``digital nudges'' can be used to steer user behavior toward socially desirable outcomes \citep{thaler2008nudge}, such as reduced digital addiction \citep{allcott2022digital}. Related analytical work studies how platforms strategically design information delivery to influence user checking and engagement behavior, including push notifications as dynamic information design \citep{iyer2022pushing}. More broadly, marketing research has a long tradition of studying the drivers of prosocial consumer behavior, such as the role of peer effects in the adoption of ``green'' technology \citep{bollinger2012peer}, the impact of resource mindsets and self-construal on food waste reduction \citep{gao2024resources}, and the framing of cause-related marketing appeals \citep{tsiros2020lowering}.

However, the path from a company ``doing good'' to its customers ``doing well'' is complex and fraught with unintended consequences. Foundational work questioned whether CSR initiatives always lead to better consumer responses, finding they can, under certain conditions, decrease purchase intentions \citep{sen2001does} or harm customer satisfaction and lower the firm's market value \citep{luo2006corporate}. These backfire effects are often attributed to psychological mechanisms. For example, environmental appeals for ``rescue-based foods'' inadvertently trigger mental imagery of waste and subsequently reduce demand \citep{de2021how}, and cause-related campaigns suggesting a high minimum donation (a seemingly more generous offer) reduce participation by diminishing the consumer's feeling of personal contribution \citep{tsiros2020lowering}.

We distinguish our work by identifying a mechanism for unintended consequences that is beyond psychological reasons. We document an ``algorithmic nudge reversal,'' where a well-being intervention explicitly designed to reduce unhealthy platform engagement paradoxically caused a sustained increase in platform usage. We show that this unintended outcome is driven by the recommender system's endogenous adaptive learning. The platform's recommendation algorithm used the nudge as ``forced exploration,'' which allows it to use the resulting data to correct embedding biases and optimize its content-matching policy. Consequently, our findings highlight that in automated environments, prosocial interventions must be evaluated not just for their behavioral impact on users, but for the training signal they provide to the underlying machine learning model.

\section{Conceptual Framework and Testable Implications}\label{sec:model}

This section lays out the mechanism that guides the empirical analysis. The formal model is placed in Appendix A. The key distinction is between a temporary intervention and the data it generates for an adaptive recommender system. The campaign boost lasts only during the intervention window, but users' responses to the boosted content can enter the platform's learning pipeline and affect the policy used after the campaign.

The framework begins with a focal content category, celebrity-related sleep-reminder videos. The recommender system has a learned representation of a user's preference for this category and chooses a recommendation policy that determines the effective intensity of focal content in the feed. If the learned representation understates the user's latent preference, the baseline policy under-supplies focal content. Routine feedback then becomes self-censoring: because the system rarely exposes the user to the focal category, the user generates little informative behavior about that category. Such selection effects, in which algorithmic exposure systematically shapes the data later used to train the system, are documented in the recommender-system and digital-platform literatures \citep{lambrecht2019algorithmic, chaney2018algorithmic}. When updating the recommendation policy is costly, discrete, or subject to retraining thresholds, these routine signals can fall inside an inaction zone. This logic parallels classical state-dependent pricing models, in which fixed adjustment costs create regions of policy inaction \citep{caplin1991state}, and is empirically consistent with evidence that high menu costs in retail pricing produce infrequent price adjustments \citep{stamatopoulos2021menu}. The system may therefore keep using the incumbent policy even when its representation is biased.

The intervention changes this information environment. At the beginning of the campaign period, the platform applies a score boost to the reminder videos for treated users during the late-night window. The boost does not force exposure; it increases the probability that reminder videos clear the ranking threshold and enter the feed. This temporary policy perturbation has two effects. First, it can directly and unintentionally improve match quality if the baseline system had under-supplied celebrity-related content. In that case, even a campaign designed to discourage usage can raise engagement because it moves the feed closer to latent demand. Second, it makes user responses more informative. By showing content that the baseline policy would have served less often, the intervention creates new behavioral data about preferences that were previously difficult for the system to learn.

The post-intervention period is the crucial step. Once the boost is removed, any continuing effect cannot be attributed to the mechanical presence of the campaign videos. Persistence arises only if the intervention-period responses are incorporated into the recommender's subsequent update. If these responses move the learned representation outside the inaction zone, the platform changes the recommendation policy used after the campaign. The long-run effect is therefore a policy-update effect: the feed changes because the system has learned from user responses to the temporary intervention.

This framework yields five testable implications, each of which is examined empirically in Sections~\ref{sec:results} and \ref{sec:mechanism}, and corresponds to a formal statement in Appendix~A.

\begin{description}
\item[Implication 1 (Manipulation Channel).] The intervention should operate through the intended recommendation channel. Treated users who are active during the late-night window should be more likely to receive reminder videos, while users who never enter that window should remain effectively unaffected. This implication concerns experimental validity rather than recommender-system learning per se.
\item[Implication 2 (Engagement During the Campaign).] If the baseline policy under-supplies focal content, the intervention can increase engagement during the campaign rather than reduce it, because the boost moves the feed closer to latent demand (formalized in the period-1 statement of Proposition~\ref{prop:2}, Appendix~A).
\item[Implication 3 (Post-Campaign Persistence).] If the intervention generates a sufficiently informative signal to move the system out of its inaction zone, engagement can remain elevated after the campaign because the post-campaign policy is closer to latent preferences (formalized in the period-2 statement of Proposition~\ref{prop:2}, building on Proposition~\ref{prop:1}, Appendix~A).
\item[Implication 4 (Cross-Category Reallocation).] The policy update should reshape the feed beyond the reminder videos themselves. Categories related to celebrity entertainment should gain exposure. In contrast, because user attention and feed slots are finite, less related categories may lose share even when aggregate engagement increases (formalized in Proposition~\ref{prop:3}, Appendix~A).
\item[Implication 5 (Heterogeneity by Pre-Campaign Exposure).] Treatment effects should be heterogeneous: the magnitude of post-campaign updating depends on the misalignment between the platform's learned representation and the user's latent preference, on the informativeness of the intervention-generated signal, and on the cost of updating (formalized in Proposition~\ref{prop:4}, Appendix~A). Users for whom the pre-campaign policy leaves more room for learning should exhibit larger responses than users for whom the system is already highly confident.
\end{description}

These implications organize the empirical analysis. We first verify the manipulation and placebo conditions (Implication~1), then estimate short-run and post-campaign engagement effects (Implications~2 and~3), and finally examine category-level reallocation, conversion-rate patterns, and heterogeneity by pre-campaign recommendation exposure (Implications~4 and~5).

\section{Experimental Design and Data}\label{sec:setting}

Testing the implications laid out in Section~\ref{sec:model} requires an empirical setting in which (i) the intervention exogenously shifts users' content exposure through the recommendation policy itself, (ii) the intervention is temporary and narrowly targeted, so that post-campaign policy updating can be distinguished from short-run persuasion or novelty effects, and (iii) randomized assignment is feasible at scale. The focal celebrity-driven sleep-reminder campaign satisfies these conditions. We describe below the platform environment, the intervention, and the experimental design that map the framework's three periods into the data: the pre-campaign baseline, the campaign week, and the post-campaign weeks.

\subsection{The Platform and Recommender System}
The experiment was conducted on one of China's largest short-video platforms. The platform boasts over 400 million daily active users and serves billions of video recommendations daily. On this platform, users act as both content consumers and creators. Similar to TikTok, the user experience is centered around an endlessly scrolling, algorithmically curated ``For You" feed of short videos, typically lasting from a few seconds to a few minutes. A typical user's feed involves approximately 400 videos daily, from which the user watches an average of 120 videos for 5 seconds or more, spending roughly 1.8 hours daily on the app.

The platform's user interface is designed for immersive and seamless content consumption. Upon launching the application, the user is immediately presented with a full-screen video that starts playing automatically. The primary mode of navigation is a vertical swipe; a user scrolls up to dismiss the current video and instantly load the next one from their personalized feed. If a user does not scroll after a video finishes, it replays automatically. This design has a key methodological implication: because videos are directly presented to the user, for \emph{every} recommended video, the user at least briefly considers whether to watch it or not. This differs from traditional recommender systems, such as Amazon, where users may easily ignore suggested items without even considering them.

Following the standard of modern content-sharing social networks, the platform's multi-faceted revenue model relies on user engagement to derive income from three main sources. First, online advertising generates revenue through impressions and clicks on ads integrated into the user's ``For You" video feed. Second, the platform collects commissions on virtual gifts sent to live-streamers by their audience. Third, it also functions as an e-commerce platform and collects sales commissions by enabling content creators to embed and sell products in their videos. As with TikTok or Instagram, these revenue streams depend heavily on the recommender system's capacity to maximize app usage, cementing aggregate engagement as the platform's central business metric.

\begin{figure}[t!]
    \centering
    \includegraphics[width=0.9\textwidth]{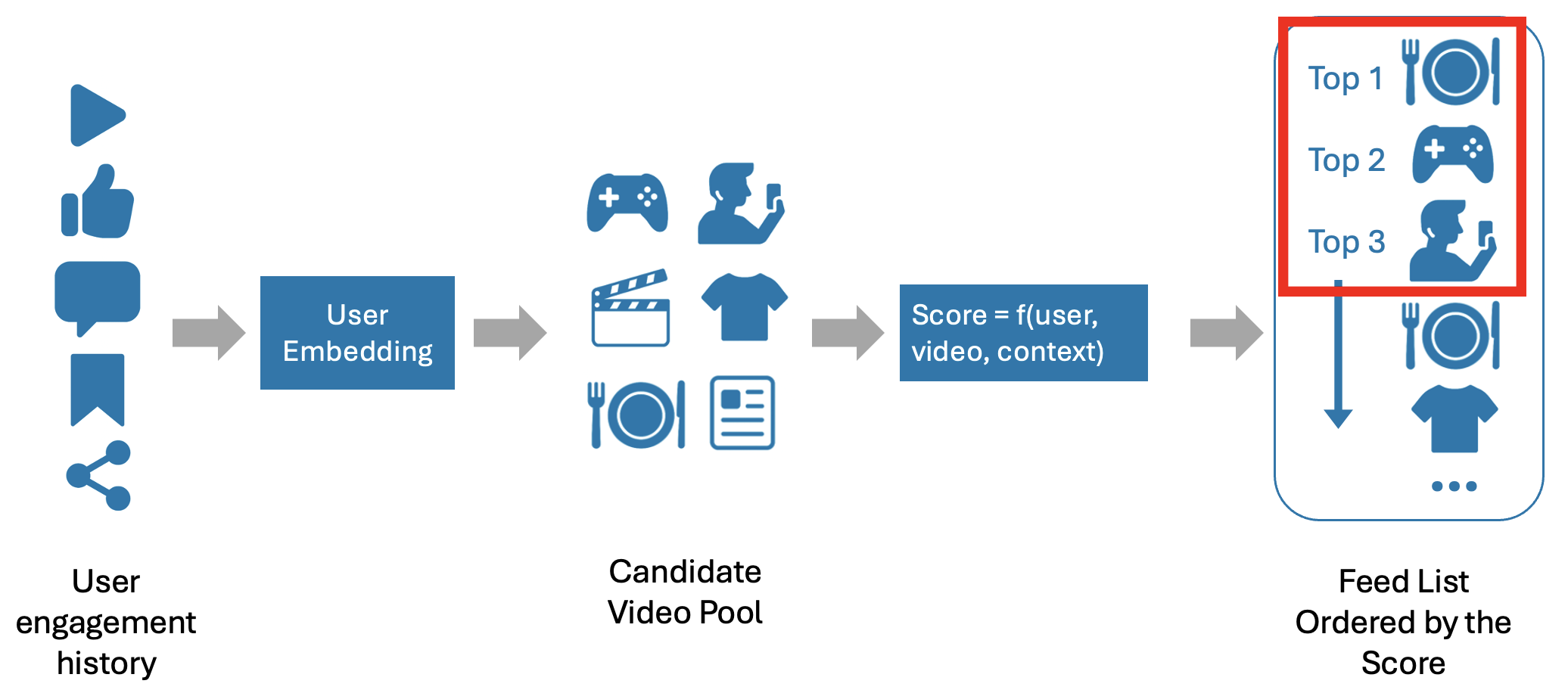}
    \caption{The Feed Generation Process in the Recommender System}
    \begin{minipage}{0.92\textwidth}
    \scriptsize
    \textit{Notes:} This figure illustrates the multi-stage feed-generation process of the recommender system, including candidate retrieval, user-video scoring based on embeddings, ranking, and final content delivery.
    \end{minipage}
    \label{fig:feed_generate}
\end{figure}

The platform’s recommender system is a sophisticated, multi-stage engine designed to generate a personalized experience, as illustrated in Figure \ref{fig:feed_generate}. When a user opens the app or swipes down the feed, a video retrieval request is sent to the server, initiating the recommendation process. The system narrows a vast pool of billions of videos into a candidate set of several million. These millions of candidate videos are then evaluated based on the interaction among video-specific features, additional contextual factors (e.g., location, season, and time of the day), and the user's embedding, a high-dimensional vector representing the user’s content preferences inferred from their past engagement patterns. Specifically, each video receives a personalized score by weighting predicted measures of various user actions, such as viewing duration, like, and share. The weights are determined by the business goals of the platform. Finally, the system ranks these candidate videos according to the personalized scores and serves approximately 10 to 20 of the highest-ranked videos from the server to the user’s device.

\subsection{The Intervention: A Celebrity-Driven Sleep Reminder Campaign}\label{subsec:intervention}

The experiment was designed as a content-level boosting intervention embedded directly within the recommender system’s ranking process. The platform collaborated with twenty-eight well-known celebrities to produce short sleep-reminder videos, each generally under thirty seconds in length. In these videos, the celebrities appeared sincere and affable, addressing viewers directly as if speaking to a friend and reminding them that it was late and time to go to sleep. Before the campaign began, users were randomly assigned to either a treatment or a control group. The treatment group's recommendation feed received a boost for the sleep-reminder content, while the control group experienced the standard recommendation policy.

From a business perspective, the campaign served a dual purpose. Publicly, it was promoted as a corporate social responsibility initiative to demonstrate the platform’s commitment to user well-being, acknowledging excessive late-night screen time can cause sleep deprivation and eventual disengagement in the long term. By encouraging healthier usage patterns, the platform aimed to project a positive brand image and address broader societal concerns about digital well-being. Internally, however, the campaign also offered an opportunity for a controlled, large-scale field experiment. Because the intervention could be precisely implemented within the recommendation architecture, it enabled the platform to study how a specific modification to the ranking rule influences both user behavior and the system’s downstream learning process.

\begin{figure}[t!]
    \centering
    \includegraphics[width=0.9\textwidth]{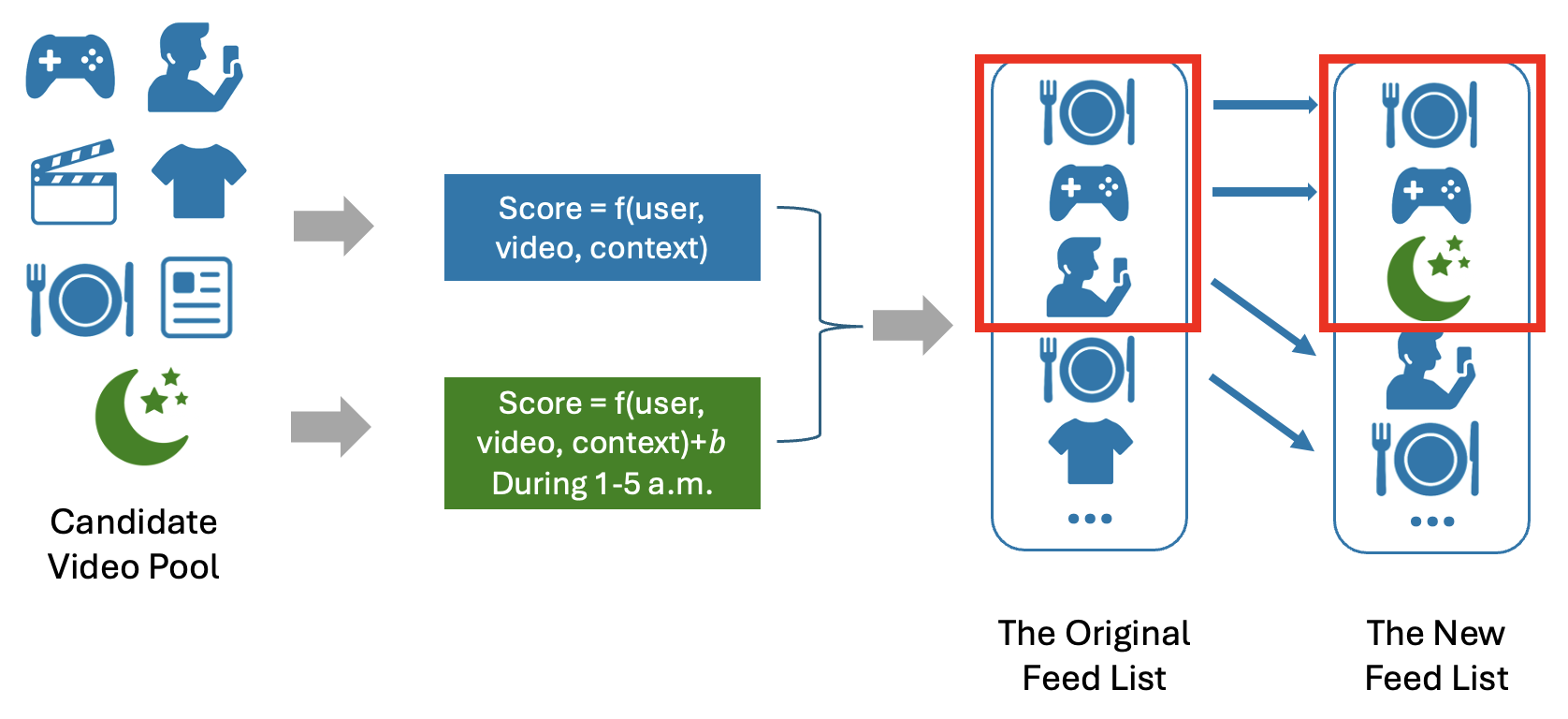}
    \caption{Mechanism of Content Boosting in the Recommender System}
    \begin{minipage}{0.92\textwidth}
    \scriptsize
    \textit{Notes:} This figure illustrates how the sleep-reminder intervention modifies the ranking stage of the recommender system by adding a small positive score adjustment to sleep reminder videos during late-night hours for treated users, increasing their probability of appearing in the feed without forcing exposure.
    \end{minipage}
    \label{fig:boost}
\end{figure}

The technical mechanism of boosting the target content is illustrated in Figure \ref{fig:boost}. As described earlier, each time a specific user requested a new feed, the recommender first retrieved a large set of candidate videos from the global content pool, including the tagged sleep-reminder clips. It then computed the personalized score of each video based on user embedding ($user$), video features ($video$), and contextual information ($context$)
\begin{equation}
Score = f(user, video, context),
\end{equation}
and the videos were ranked by their scores to form the personalized feed shown to the user.\footnote{To minimize cluttering, we suppress the subscripts.} For the control group and for treated users outside the late-night period, this standard scoring logic remained unchanged.

The only modification during this recommendation process occurred for treated users during the late-night window from 1:00 a.m. to 5:00 a.m. During those hours, the system applied a small positive adjustment to treated users' personalized scores of any sleep-reminder videos:
\begin{equation}
Score = f(user, video, context) + b \cdot \mathds{1}\{\text{treated user, late-night window, no prior reminder exposure}\},
\end{equation}
where $b$ is a positive constant determined by the platform.\footnote{The platform does not reveal the level of $b$ to us.} This additional term functioned as a subtle but systematic boost to the scores of the reminder videos during the ranking stage. After the modified scores were computed, the candidate videos were ranked, and the top 10 to 20 items were served to the user’s device in the same way as under the standard policy.

This simple change produced a clear mechanical consequence in the ranking outcome. In the baseline scenario with the standard policy, a sleep-reminder video might have ranked too low to be served to the user, being crowded out by videos with higher scores. Under the treatment condition, the boost factor could be sufficient to lift the reminder video's score above the ranking threshold, pushing it into the user’s feed by displacing another video that would otherwise have appeared. Therefore, the intervention did not mandate an exposure; instead, it increased the reminder video's score and subsequent probability of being served to the treated user. The boost was also applied only until a reminder video was first served: once a reminder video appeared in a treated user's feed, the score adjustment was switched off for that user, so each treated user received at most one boosted exposure rather than repeated boosting.

The design maintained the same retrieval and ranking logic for all other content, ensuring that the only experimental variation arose from the additive term $b$. This structure enabled clean identification of the causal effects of content boosting on both user behavior and the platform’s subsequent recommendation adjustments. As the next section shows, the manipulation effectively increased the late-night exposure and viewing of the celebrity reminder videos. More importantly, through the system’s continuous learning and embedding updates, this targeted intervention triggered lasting changes in the composition of users’ feeds and in their overall engagement patterns.

\subsection{Sample, Data, and Randomization Check}
The experiment ran from April 29 to May 6, 2022. The platform randomly assigned 10\% of its registered users to the treatment group and 90\% to the control group. From this large-scale experiment, we randomly sampled a balanced panel of 200,000 users who were active on the first day of the experiment, with 100,000 users drawn from the treatment group and 100,000 from the control group. This subsample underpins our main analyses. We tracked these users' user-level demographics (e.g., gender, city tier) and engagement metrics (e.g., app usage, video duration) across the pre-experimental, experimental, and post-experimental periods.

Table \ref{tab:celebrities_info_sorted} shows that the participating celebrities of the reminder videos span a wide range of profiles. There are twelve women and sixteen men. Ages range from early twenties to early fifties. The population of fan bases varies substantially, from about 100,000 to more than 10 million, with a median of 1.12 million and an average of 3.28 million. The celebrities' occupations cover actors, singers, hosts, comedians, dancers, a pianist, and a photographer. All ages and fan counts refer to the first day of the treatment period. This breadth means the boosted clips were supplied by figures who speak to very different audiences and who enter the ranking list with very different levels of organic strength. The intervention, therefore, operated on reminder videos with diverse baseline appeal. 

\begin{table}[t!]
    \centering
    \caption{Participating Celebrities and Basic Characteristics}
    \label{tab:celebrities_info_sorted}
    \begin{threeparttable}
    \scriptsize
    \begin{tabular}{lccc}
        \toprule
        Age & Gender & Fan Base Size & Occupation \\
        \midrule
        47 & M & 10m+     & Actor, Director \\
        37 & M & 10m+     & Comedian, Actor \\
        37 & F & 10m+     & Actress, Host, Singer \\
        29 & M & 5--10m   & Actor \\
        41 & M & 5--10m   & Host \\
        31 & F & 5--10m   & Actress \\
        40 & M & 1--5m    & Actor \\
        39 & M & 1--5m    & Pianist \\
        43 & M & 1--5m    & Actor \\
        28 & F & 1--5m    & Actress \\
        24 & F & 1--5m    & Singer, Dancer \\
        41 & M & 1--5m    & Actor, Singer \\
        35 & F & 1--5m    & Actress, Photographer \\
        43 & F & 1--5m    & Actress \\
        40 & F & 100k--1m & Actress, Director \\
        23 & F & 100k--1m & Singer \\
        40 & M & 100k--1m & Comedian, Actor \\
        52 & M & 100k--1m & Singer, Host \\
        26 & M & 100k--1m & Singer, Host \\
        24 & F & 100k--1m & Singer \\
        25 & F & 100k--1m & Singer \\
        27 & M & 100k--1m & Actor \\
        50 & M & 100k--1m & Singer \\
        33 & F & 100k--1m & Singer, Actress \\
        41 & M & 100k--1m & Actor \\
        24 & F & 100k--1m & Singer \\
        37 & M & 100k--1m & Actor \\
        24 & M & 100k--1m & Comedian, Actor \\
        \bottomrule
    \end{tabular}
    \begin{tablenotes}
    \scriptsize
    \item \textit{Notes:} This table reports basic characteristics of the celebrities who participated in the sleep-reminder campaign. Each celebrity recorded short reminder videos that were eligible for temporary ranking boosts during late-night hours for users assigned to the treatment group. Age and fan base size are measured as of the first day of the experimental period. Fan base size is defined as the number of followers on the platform and is discretized into four categories for confidentiality and interpretability: 100k--1m (100,000 to 1 million followers), 1--5m (1 to 5 million), 5--10m (5 to 10 million), and 10m+ (more than 10 million followers). Occupation categories reflect publicly recognized professional roles on the platform and are not mutually exclusive. The table is sorted in descending order of fan base size.
    \end{tablenotes}
    \end{threeparttable}
\end{table}

Table \ref{tab:main_variables} provides descriptions for the main variables used in our empirical analysis. Our key outcome variables are measured at the user-day level. Note that, for confidentiality reason, the labels of most demographic sub-categories have been anonymized. For the same reason, we cannot show the summary statistics of certain variables. App Usage is the total time in hours a user spent on the app on a given day. Night Usage is the time in hours spent between 1 AM and 5 AM. Video Duration is the total duration in hours of all videos played by the user. We also analyze the composition of a user's feed, using measures such as the Recommended Ratio and Played Ratio for various content categories, which represent the proportion of videos from a given category in the user's recommended and actually played content, respectively.

\begin{table}[t!]
\centering
\caption{Descriptions of Main Variables Used in the Empirical Analysis}
\label{tab:main_variables}
\scriptsize
\begin{tabular}{p{4cm} p{7cm} p{3cm}}
\toprule
\textbf{Variable Name} & \textbf{Description} & \textbf{Granularity} \\
\midrule
\multicolumn{3}{l}{\textbf{Outcome Variables}} \\
App Usage & Total time (in hours) spent on the app. & User-Day \\
Night Usage & Total time (in hours) spent on the app between 1 AM and 5 AM. & User-Day \\
Video Duration & Total duration (in hours) of all videos played. & User-Day \\
Live Duration & Total time (in hours) spent watching live-stream broadcasts. & User-Day \\
Recommended/Played Ratio & Share (\%) of videos from a specific content category in the user's recommended/played content. & User-Day-Category \\
Recommended/Played Count & Absolute number of videos from a specific content category in the user's recommended/played content. & User-Day-Category \\
Celebrity Content Ratio & Share (\%) of celebrity entertainment videos among all videos played. & User-Day \\
\midrule
\multicolumn{3}{l}{\textbf{Demographic Variables}} \\
Age & User's age in years. & User \\
Gender & User's gender (Female, Male, Unknown). & User \\
Community Type & Classification of user's location (e.g., Urban, Rural). & User \\
City Level & Tier of the user's city (e.g., Tier 1, Tier 2). & User \\
Price Range & Price range of the user's mobile device. & User \\

\bottomrule
\end{tabular}
\end{table}

To verify the integrity of the experiment's random assignment, we conducted a comprehensive balance check on pre-treatment variables. As shown in Table \ref{tab:randomization_check}, we find no significant differences between the treatment and control groups across both demographic characteristics and key behavioral metrics from the week prior to the intervention. Panel A shows that demographic variables, including gender, community type, city level, device price range, and age, are well-balanced, with negligible absolute differences in proportions and a statistically insignificant difference in mean age. Panel B confirms that pre-treatment engagement patterns are also balanced. Between treatment and control groups, the standardized differences for app usage, night usage, video duration, and live duration are all minimal (less than 0.01 in magnitude), and t-tests fail to reject the null hypothesis of these variables having equal means between groups. This balance check provides strong evidence that the randomization was successful, enabling us to attribute any post-treatment differences to the causal effect of the intervention.

\begin{table}[t!]
\centering
\scriptsize
\caption{Randomization Check on Pre-Treatment Characteristics}
\label{tab:randomization_check}
\begin{threeparttable}

\begin{tabular*}{\linewidth}{@{\extracolsep{\fill}}lccc}
\toprule
\multicolumn{4}{c}{\textbf{Panel A: Demographics}} \\
\midrule
Variable & Chi-square & p-value & Max Abs. Diff. (\%) \\
\midrule
Gender & 0.365 & 0.833 & 0.040 \\
Community Type & 2.000 & 0.736 & 0.163 \\
City Level & 6.501 & 0.369 & 0.341 \\
Device Price Range & 2.871 & 0.720 & 0.228 \\
\bottomrule
\end{tabular*}

\vspace{0.8em}

\begin{tabular*}{\linewidth}{@{\extracolsep{\fill}}lccc}
\toprule
\multicolumn{4}{c}{\textbf{Panel B: Metrics in One Week Prior to the Experiment}} \\
\midrule
Variable & Treatment Mean & Control Mean & p-value \\
\midrule
App Usage & 0.9892 & 0.9904 & 0.7838 \\
Night Usage & 0.4905 & 0.4921 & 0.7269 \\
Video Duration & 0.9839 & 0.9849 & 0.8314 \\
Live Duration & 0.3423 & 0.3424 & 0.9836 \\
\bottomrule
\end{tabular*}

\begin{tablenotes}[flushleft]
\scriptsize
\item \textit{Notes:} This table reports balance checks for pre-treatment characteristics between treatment and control users.
Panel A reports tests of equality for categorical demographic variables based on chi-squared tests of independence comparing the full category distributions across groups.
“Max Abs. Diff.” reports the maximum absolute difference (in percentage points) across all categories within a variable.
Panel B reports balance for pre-treatment engagement metrics measured during the one-week period prior to the experiment; all variables are standardized to have unit standard deviation in the full sample.
P-values in Panel B are from two-sample t-tests.
None of the variables in either panel exhibit statistically significant differences between treatment and control groups.
\end{tablenotes}

\end{threeparttable}
\end{table}

\section{Empirical Analyses of the Experiment}\label{sec:results}
In this section, we estimate the causal effects of the sleep-reminder intervention on user engagement and trace their evolution over time. Motivated by the framework in Section~\ref{sec:model}, we organize the analysis around three questions.

First, did the intervention operate through the intended algorithmic channel, selectively increasing exposure to reminder content without affecting users outside the late-night window? Second, how did the intervention affect overall and late-night engagement during the campaign week? Third, do these effects persist after the intervention ends, consistent with recommender-system updating rather than a transitory response to the reminder videos?

We address these questions sequentially, moving from validation of the experimental manipulation to short-run treatment effects, and then to longer-run outcomes. Our analysis primarily relies on an Ordinary Least Squares (OLS) specification to estimate intent-to-treat (ITT) effects at the user-day level. Specifically, our analyses are based on the following general structure:
\begin{equation}
\text{Outcome}_{it} = \beta_0 + \beta_1 \text{Treatment}_i + \epsilon_{it}
\end{equation}
where $\text{Outcome}_{it}$ is the outcome of interest for user $i$ on day $t$, and $\text{Treatment}_i$ is a binary indicator for assignment to the treatment group. We also explore heterogeneity in the impact of the treatment along various categories. To this end, we adapt the regression framework to the user-day-category level.

To estimate the causal effect for users whose exposure status was changed by assignment, we also employ a two-stage least squares (2SLS) instrumental-variables approach and report local average treatment effects (LATE) for compliers. In this specification, random assignment ($\text{Treatment}_i$) instruments for actual exposure to a reminder video. Any variations from these core specifications are described as they appear. All regressions use robust standard errors.

\subsection{Manipulation Check}

\begin{table}[t!]
\centering
\scriptsize
\caption{Manipulation Check}
\label{tab:manipulation_check}
\begin{threeparttable}

\begin{tabular}{lcc}
\toprule
 & \textbf{Never-Night Users} & \textbf{Full Sample} \\
 & (1) & (2) \\
\midrule
\multicolumn{3}{l}{\textbf{Dependent Variable: Treated (0/1)}} \\
\midrule
Treatment & 0.0009 & 0.1622$^{***}$ \\
          & (0.0010) & (0.0013) \\
Constant  & 0.0178$^{***}$ & 0.0226$^{***}$ \\
          & (0.0007) & (0.0005) \\
\midrule
Control Treated Rate & 0.0178 & 0.0226 \\
Treatment Treated Rate & 0.0187 & 0.1848 \\
\midrule
Observations & 68{,}520 & 200{,}000 \\
R-squared & 0.0000 & 0.0708 \\
\bottomrule
\end{tabular}

\begin{tablenotes}
\scriptsize
\item \textit{Notes:} This table examines whether the treatment assignment (randomization) properly predicts actual treatment receipt for different user subgroups. The dependent variable is a binary indicator for whether the user actually had the night reminder video in their feed (\textit{Treated}). The key independent variable is treatment assignment (\textit{Treatment}). ``Never-Night Users'' are defined as users who had zero night usage (night\_usage = 0) on all days during the experimental period (post=1), indicating they never used the app during late-night hours throughout the experiment. These users should not be affected by the late-night intervention. Column (1) shows results for never-night users only, while Column (2) shows results for the full sample. The analysis tests whether the relationship between treatment assignment and actual treatment receipt differs for users who never engaged in night usage during the experiment. The ``Control Treated Rate'' and ``Treatment Treated Rate'' show the proportions of users exposed to the reminder videos in the control and treatment groups, respectively. Robust standard errors are in parentheses. $^{***}p<0.01$, $^{**}p<0.05$, $^{*}p<0.1$.
\end{tablenotes}
\end{threeparttable}
\end{table}

Before turning to the main results, we first verify that the intervention successfully operated through the intended channel and did not inadvertently alter other aspects of the recommendation process. The treatment directly boosted the ranking scores of sleep-reminder videos recorded by participating celebrities, with the goal of making them more likely to appear in late-night video feeds. Consistent with this design, random assignment strongly predicts actual treatment receipt in the
full sample. In Table~\ref{tab:manipulation_check}, ``Control Treated Rate" and ``Treatment Treated Rate" represent the proportions of users exposed to the sleep reminder videos in the control and treatment groups, respectively. Column (2) of the table shows that 18.48\% of users in the treatment group received at least one reminder video in their video feeds, but only 2.26\% of the control group received the video, a stark 16.22\% difference. The coefficient estimate for ``Treatment" is statistically significant, and the effect size corroborates the difference between `Control Treated Rate" and ``Treatment Treated Rate". This confirms that the manipulation successfully increased exposure to the targeted content among users assigned to the treatment.

Note that the intervention was only activated when users opened the app between 1 a.m.\ and 5 a.m.\ during the treatment period. We label those who never opened the app during these hours throughout the experiment as ``Never-night'' users. These users should therefore remain unaffected even if they were assigned to the treatment group. Column (1) of Table~\ref{tab:manipulation_check} confirms this intuition: the coefficient for \textit{Treatment} is statistically insignificant (0.0009 with s.e.\ 0.0010), and the proportions of treatment group and control group users exposed to the reminder videos are almost identical. In the spirit of placebo tests, Table \ref{tab:placebo_never_night} further shows that for this same never-night subgroup, treatment and control users exhibit no statistically significant differences in daily App Usage and Video Duration during the experiment period, confirming that the late-night boosting rule was implemented correctly and did not affect users outside its intended scope.\footnote{Night Usage is omitted from this placebo test because the never-night subgroup is defined by having zero night usage throughout the experiment, making the outcome mechanically zero. In Appendix Table~\ref{tab:placebo_never_night_preexp}, we report an alternative version of this placebo test in which never-nighters are defined based on pre-experiment behavior rather than during-experiment behavior. The results are qualitatively similar.} Together, these results show that the manipulation operated as designed: users who should be affected are more likely to receive the reminder videos, while users outside the intended late-night window remain effectively untreated.

\begin{table}[t!]
\centering
\scriptsize
\caption{Placebo Test: Treatment Effects on Never-Night Users}
\label{tab:placebo_never_night}
\begin{threeparttable}
\begin{tabular}{lcc}
\toprule
& App Usage & Video Duration \\
\midrule
Treatment & 0.0013 & 0.0003 \\
& (0.0014) & (0.0016) \\
Constant & 0.5959$^{***}$ & 0.5650$^{***}$ \\
& (0.0009) & (0.0010) \\
\midrule
Observations & 548,160 & 548,160 \\
\bottomrule
\end{tabular}
\begin{tablenotes}
\scriptsize \item \textit{Notes:} This table presents the results of OLS regressions analyzing the treatment effect on users who had zero night usage (night\_usage = 0) on all days during the experimental period, i.e., users who never used the app during late-night hours throughout the experiment. This definition is consistent with the ``Never-Night Users'' in the manipulation check (Table~\ref{tab:manipulation_check}). Because these users never opened the app during the late-night window when the intervention was active, treatment assignment should have no effect on this subgroup. Night Usage is omitted as a dependent variable because it is mechanically zero for this sample by construction. The absence of statistically significant effects on App Usage and Video Duration confirms that the intervention did not spill over to users outside its intended scope. Dependent variables are standardized to have unit standard deviation. Robust standard errors are in parentheses. $^{*}$ p $<$ 0.1, $^{**}$ p $<$ 0.05, $^{***}$ p $<$ 0.01.
\end{tablenotes}
\end{threeparttable}
\end{table}

\subsection{Treatment Effects on Usage}

We next evaluate how the intervention affected overall engagement during and after the experimental period. As discussed in Section~\ref{subsec:intervention}, assignment to treatment makes exposure to a sleep-reminder video more likely, but exposure is not guaranteed because the final feed still depends on the ranking procedure. Accordingly, we report both intent-to-treat effects (ITT) and local average treatment effects (LATE). Table~\ref{tab:shortterm_effects} reports short-term effects during the campaign week (Week 0). Panel A presents ITT estimates from OLS regressions of daily outcomes on treatment assignment, and Panel B reports LATE estimates from two-stage least squares regressions using treatment assignment as an instrument for actual exposure. We report heteroskedasticity-robust standard errors and compute relative effects as the coefficient divided by the corresponding control-group mean. 

The results indicate economically and statistically significant increases in engagement during the experiment week. In terms of intent-to-treat effects on user engagement metrics (Panel A), the treatment group demonstrates significantly higher levels than the control group in their App Usage, Night Usage, and Video Duration (Columns 1-3). In relative terms, the treatment group's App Usage, Night Usage, and Video Duration are 2.18\%, 14.75\%, and 2.71\% higher than the control group's metrics. Panel B further shows that, for the compliers who are actually exposed to reminder videos, the implied LATE are much larger: App Usage rises by 13.51\%, Night Usage by 92.85\%, and video duration by 16.77\%. Taken together, the estimates show that a seemingly small perturbation to late-night recommendations leads to a broad and meaningful increase in short-video consumption rather than a simple substitution across content types.

As a placebo check, Column 4 shows the treatment effects on the engagement of live-stream viewing. Live-stream viewing involves users watching streamers' live broadcasting and is outside the short-video feed-ranking intervention. Accordingly, the intervention should not affect this engagement margin among the treatment group. As expected, Column 4 shows statistically insignificant ITT and LATE for this type of engagement.

\begin{table}[t!]
\centering
\scriptsize
\caption{Short-Term Treatment Effects During the Experimental Week}
\label{tab:shortterm_effects}
\begin{threeparttable}

\begin{tabular}{lcccc}
\toprule
\multicolumn{5}{l}{\textbf{Panel A: OLS (ITT, DV $\sim$ Treatment)}} \\
\midrule
 & \textbf{App Usage} & \textbf{Night Usage} & \textbf{Video Duration} & \textbf{Live Duration} \\
 & (1) & (2) & (3) & (4) \\
\cmidrule(lr){2-5}
Treatment & 0.0202$^{***}$ & 0.0574$^{***}$ & 0.0238$^{***}$ & 0.0002 \\
          & (0.0016) & (0.0016) & (0.0016) & (0.0016) \\
Constant  & 0.9240$^{***}$ & 0.3893$^{***}$ & 0.8781$^{***}$ & 0.3123$^{***}$ \\
          & (0.0011) & (0.0011) & (0.0011) & (0.0011) \\
\midrule
Relative Effect (ITT, \%) & 2.18 & 14.75 & 2.71 & -- \\
Observations & 1{,}600{,}000 & 1{,}600{,}000 & 1{,}600{,}000 & 1{,}600{,}000 \\
\bottomrule
\end{tabular}

\vspace{0.8em}

\begin{tabular}{lcccc}
\toprule
\multicolumn{5}{l}{\textbf{Panel B: IV (LATE, DV $\sim$ Treated)}} \\
\midrule
 & \textbf{App Usage} & \textbf{Night Usage} & \textbf{Video Duration} & \textbf{Live Duration} \\
\cmidrule(lr){2-5}
Treated   & 0.1244$^{***}$ & 0.3540$^{***}$ & 0.1467$^{***}$ & 0.0013 \\
          & (0.0097) & (0.0095) & (0.0096) & (0.0097) \\
Constant  & 0.9212$^{***}$ & 0.3813$^{***}$ & 0.8748$^{***}$ & 0.3122$^{***}$ \\
          & (0.0013) & (0.0012) & (0.0013) & (0.0013) \\
\midrule
\multicolumn{5}{l}{\textit{First Stage: Treated $\sim$ Treatment}} \\
Treatment & 0.1622$^{***}$ & 0.1622$^{***}$ & 0.1622$^{***}$ & 0.1622$^{***}$ \\
          & (0.0005) & (0.0005) & (0.0005) & (0.0005) \\
First-stage F-stat & 121820.65 & 121820.65 & 121820.65 & 121820.65 \\
\midrule
Relative Effect (LATE, \%) & 13.51 & 92.85 & 16.77 & -- \\
Observations & 1{,}600{,}000 & 1{,}600{,}000 & 1{,}600{,}000 & 1{,}600{,}000 \\
\bottomrule
\end{tabular}

\begin{tablenotes}
\scriptsize
\item \textit{Notes:} This table reports treatment effects on user engagement during the experimental week (Week 0). The analysis is at the user-day level. Dependent variables are standardized to have unit standard deviation. Panel A reports the intent-to-treat (ITT) effect from OLS regressions of each outcome on the treatment assignment dummy. Panel B reports the local average treatment effect (LATE) from a 2SLS regression, where random assignment to the treatment group is used as an instrumental variable for whether the user was actually exposed to a reminder video. The first stage shows the effect of treatment assignment on actual treatment receipt. Robust standard errors are in parentheses. $^{***}p<0.01$, $^{**}p<0.05$, $^{*}p<0.1$. Relative effects are computed as 100$\times$Coefficient / Constant and are omitted for non-significant coefficients (p $\geq$ 0.1).
\end{tablenotes}
\end{threeparttable}
\end{table}

To assess persistence of these effects, we estimate analogous specifications at the user-week level for the nine weeks starting with the campaign week (Week 0). Appendix Tables~\ref{tab:longterm_appusage} through \ref{tab:longterm_liveduration} report the long-term treatment effects for total app usage, night usage, video duration, and live duration. For total app usage, the ITT effect is about 2.1\% in Week 0 and stabilizes around 0.8 to 1.0\% in Weeks 1 to 6 before tapering off and becoming statistically indistinguishable from zero thereafter. Night usage exhibits much larger and more persistent effects: the ITT impact is about 13.8\% in Week 0 and remains between roughly 3.6\% and 6.9\% through Week 8. The IV estimates are correspondingly larger, with late-night usage for compliers remaining about 22 to 43\% higher than the control mean several weeks after the campaign. Video duration shows a similar, though slightly smaller, pattern of sustained gains, while live-stream viewing remains effectively indistinguishable from the control group throughout because it falls outside the targeted intervention. Taken together, the short-run and long-run estimates reveal a sizable and long-lasting increase in short-video engagement, concentrated in late-night consumption.

How large are these effects relative to prior platform experiments? On our partner platform, the average user is served roughly 400 videos per day, so the campaign altered the ranking of only at most one in 400 recommendations (around 0.25\%) for treated users. Yet this tiny, localized change generates an ITT effect increase of about 2\% in total usage and nearly 15\% in night usage during the campaign week, with positive effects on engagement persisting for multiple weeks. Although direct comparisons across settings are imperfect given differences in outcome definitions, study populations, and platform mechanics, the magnitudes appear unusually large relative to other randomized interventions on major social media platforms given how little of the feed is directly manipulated. For example, \citet{eckles2016estimating} find that peer encouragement on Facebook increases posting by about 0.7\% and downstream engagement by 1 to 10\%, while \citet{allcott2020welfare} and \citet{guess2023} show that much more drastic interventions, such as deactivating Facebook entirely or switching from an algorithmic to a chronological content feed, reduce usage and engagement on the order of 20 to 50\%. Our intervention is far more subtle than shutting off the platform or overhauling the content feed, yet it produces engagement gains broadly comparable in scale to many direct behavioral nudges.

These patterns suggest that something beyond a simple, one-time content change is at work. Namely, the recommender system itself adapts to the intervention. In the next section, we examine how the campaign reshapes the content mix and user-content matching.

\section{Mechanisms and Interpretation}
\label{sec:mechanism}

The empirical results in Section~\ref{sec:results} document a persistent increase in user engagement following a short-lived and narrowly targeted content intervention. This section examines the mechanism behind that pattern. The evidence is organized around the framework in Section~\ref{sec:model} and the formal model in Appendix A: the intervention changes exposure during the campaign week, users' responses generate new training signals, and the recommender system can use those signals to update the post-campaign policy.

The analysis focuses on four empirical signatures. Implication 2 (Proposition~\ref{prop:2}) links the policy change to higher engagement during the campaign when the baseline policy under-supplies the focal category. Implication 3 (Proposition~\ref{prop:2}, building on the inaction-zone logic of Proposition~\ref{prop:1}) shows why this engagement can persist after the campaign: routine feedback may leave the incumbent policy unchanged, while intervention-generated feedback can move the system outside the update threshold. Implication 4 (Proposition~\ref{prop:3}) shows why finite attention implies category-level reallocation rather than a uniform increase in all content categories. Implication 5 (Proposition~\ref{prop:4}) predicts that the magnitude of updating should vary with the initial misalignment, signal precision, and the update threshold. We evaluate these implications in turn.

\subsection{Policy Updating and Cross-Category Spillovers}
\label{subsec:policy_update}

The framework implies that policy updating should be visible in the composition of the feed. Before the campaign, the recommender uses an incumbent policy based on its existing representation of user preferences. During the campaign, the score boost perturbs exposure and generates new responses. If those responses are incorporated into the next update, the post-campaign policy should differ from the pre-campaign policy even after the boost is removed.

Table~\ref{tab:category_ratio} provides evidence of such policy updating. Although the intervention directly targeted celebrity-recorded sleep-reminder videos, we observe statistically significant and persistent changes in recommendation shares across a wide range of content categories. These changes are not uniform. Categories that are semantically or behaviorally related to celebrity entertainment, such as Games, Anime/ACG, and Sports, experience increases in both recommended and played shares, while less related categories experience relative declines.

\begin{table}[t!]
    \centering
    \scriptsize
    \caption{Relative Effect Sizes of Treatment on Recommended and Played Ratios (\%)}
    \label{tab:category_ratio}
    \begin{threeparttable}
    \begin{tabular}{l | c c c | c c c}
        \toprule
        \textbf{Category} 
        & \multicolumn{3}{c|}{\textbf{Recommended Ratio}} 
        & \multicolumn{3}{c}{\textbf{Played Ratio}} \\
        \cmidrule(lr){2-4} \cmidrule(lr){5-7}
        & \textbf{Baseline (\%)} 
        & \multicolumn{2}{c|}{\textbf{Relative Effect (\%)}} 
        & \textbf{Baseline (\%)} 
        & \multicolumn{2}{c}{\textbf{Relative Effect (\%)}} \\
        \cmidrule(lr){3-4} \cmidrule(lr){6-7}
        & 
        & \textbf{Week 0} 
        & \textbf{Week 1--8} 
        & 
        & \textbf{Week 0} 
        & \textbf{Week 1--8} \\
        \midrule
        Games & 3.92 & 3.85*** & 3.89*** & 3.99 & 4.32*** & 4.32*** \\
        Astrology \& Fortune-telling & 0.18 & 2.95*** & 3.03*** & 0.18 & 3.85*** & 2.50*** \\
        Anime / ACG & 3.19 & 2.19*** & 2.45*** & 3.11 & 2.25*** & 2.52*** \\
        Celebrity Entertainment & 2.03 & 1.46*** & 0.58*** & 2.11 & 1.56*** & 0.96*** \\
        Sports & 1.83 & 1.20*** & 1.34*** & 2.09 & 1.76*** & 1.75*** \\
        Appearance / Looks & 1.98 & 1.11*** & 1.63*** & 1.38 & 1.47*** & 1.78*** \\
        Photography & 0.61 & 0.96** & 0.53*** & 0.52 & 0.87 & -0.10 \\
        Emotions / Relationships & 3.81 & 0.79*** & 1.02*** & 3.75 & 0.61** & 1.06*** \\
        Oddities \& Wonders & 0.50 & 0.67 & 0.02 & 0.60 & 0.04 & 0.16 \\
        Beauty / Makeup & 2.08 & 0.59* & 0.49*** & 1.66 & 0.44 & 0.35** \\
        Selfies & 6.41 & 0.45** & 0.27*** & 4.66 & 0.84*** & 0.10 \\
        Film \& TV Shows & 9.63 & 0.31 & 0.00 & 10.89 & 0.53** & 0.12 \\
        Short Dramas & 4.29 & 0.27 & 0.23* & 4.61 & 0.09 & 0.38*** \\
        Lifestyle & 2.63 & 0.23 & -0.09 & 2.62 & -0.20 & -0.18 \\
        Comedy & 4.96 & 0.07 & 0.13* & 6.06 & 0.28 & 0.47*** \\
        Science \& Law & 0.62 & 0.01 & -0.61*** & 0.69 & -0.42 & -0.71*** \\
        Education & 0.79 & -0.26 & -0.00 & 0.72 & -1.23 & -0.84*** \\
        Finance \& Business & 0.75 & -0.28 & -0.16 & 0.73 & -0.56 & -0.26 \\
        Others & 5.31 & -0.32* & -0.39*** & 3.82 & -0.71** & -0.72*** \\
        Humanities & 0.77 & -0.40 & -0.75*** & 0.81 & -0.44 & -0.66*** \\
        Travel & 0.97 & -0.41 & -0.52*** & 0.86 & -0.23 & -0.44** \\
        Pets & 1.05 & -0.55 & -0.23 & 1.13 & -0.04 & 0.30 \\
        Health & 1.99 & -0.57 & 0.45*** & 1.79 & 0.27 & 0.59*** \\
        Food & 5.07 & -0.64*** & -0.51*** & 4.84 & -0.52* & -0.39*** \\
        Fashion & 5.03 & -0.68** & -0.61*** & 3.85 & -1.26*** & -0.74*** \\
        Parenting \& Family & 2.46 & -0.87*** & -0.49*** & 2.56 & -0.99*** & -0.49*** \\
        Tech \& Digital Products & 1.69 & -0.91** & -0.24 & 1.86 & -1.95*** & -1.58*** \\
        Fitness & 0.38 & -1.01 & 0.79** & 0.34 & -0.54 & 0.59 \\
        Civic News & 4.85 & -1.14*** & -0.94*** & 6.14 & -0.95*** & -0.74*** \\
        Rural / Agriculture & 3.25 & -1.21*** & -1.47*** & 3.14 & -1.29*** & -1.48*** \\
        Music & 3.80 & -1.26*** & -1.00*** & 3.36 & -1.28*** & -1.08*** \\
        Real Estate \& Home & 2.56 & -1.30*** & -1.18*** & 2.60 & -0.89** & -1.13*** \\
        Art & 1.63 & -1.30*** & -0.89*** & 1.62 & -0.59 & -0.29* \\
        Dance & 1.90 & -1.38*** & -0.72*** & 1.68 & -1.98*** & -1.01*** \\
        Military & 1.02 & -1.45*** & -1.32*** & 1.18 & -1.65*** & -1.08*** \\
        Automobiles & 2.33 & -1.47*** & -1.79*** & 2.24 & -1.46*** & -1.93*** \\
        Current Affairs \& News & 0.61 & -1.83*** & -1.29*** & 0.70 & -2.11*** & -1.16*** \\
        Reading / Books & 1.26 & -2.42*** & -2.41*** & 1.23 & -1.94*** & -2.37*** \\
        \bottomrule
    \end{tabular}
    \begin{tablenotes}
    \scriptsize
    \item Note: This table shows the treatment's effects on the share of different content categories. The analysis is at the user-day-category level. ``Baseline (\%)" columns show the average share of videos from each category for the control group. ``Relative Effect (\%)" columns report the ITT effect as a percentage, calculated from our main OLS specification as (coefficient on Treatment / baseline) $\times$ 100. The effects are shown separately for the experimental week (Week 0) and the following eight weeks (Week 1-8). Positive values indicate an increase in category share, while negative values indicate substitution away from the category.$^{*}$p$<$0.1; $^{**}$p$<$0.05; $^{***}$p$<$0.01.
    \end{tablenotes}
    \end{threeparttable}
\end{table}

\begin{figure}[t!]
    \centering
    \includegraphics[width=0.7\textwidth]{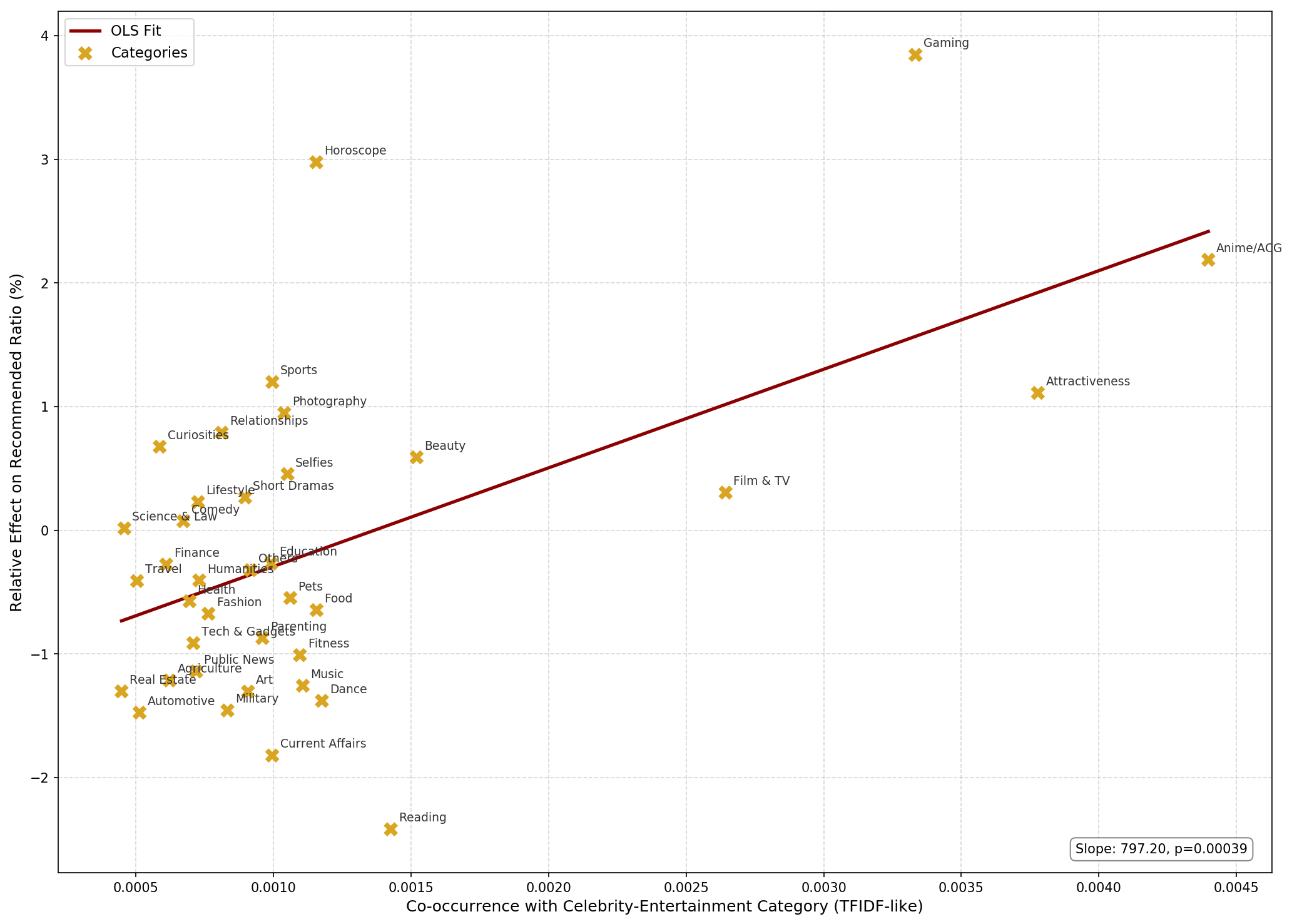}
    \caption{Spillover Effect vs. Content Co-occurrence}
    \label{fig:cooccurrence_rec_ratio}
    \begin{minipage}{0.9\textwidth}
        \scriptsize \textit{Notes:} Each point represents a content category. The y-axis shows the relative treatment effect on the recommended ratio for each category. The x-axis is a TF-IDF-like measure of the category's co-occurrence with the Celebrity-Entertainment category in the feeds of high-affinity users pre-experiment. The positive slope (797.20, p=0.00039) indicates that categories more related to celebrity content experienced larger recommendation increases following the intervention.
    \end{minipage}
\end{figure}

To further assess this mechanism, we construct a measure of content similarity based on user behavior before the experiment: a TF-IDF-like co-occurrence score between each category and the ``Celebrity Entertainment'' category. This score is higher for categories that appear more frequently in the feeds of users with stronger baseline affinity for celebrity content. Figure~\ref{fig:cooccurrence_rec_ratio} plots the relative effect on a category's recommendation ratio against this co-occurrence measure. The relationship is strong and statistically significant (slope = 797.20, $p < 0.001$), indicating that the platform's policy adjustment disproportionately favored categories related to the focal category of the intervention. Related work shows that platform-level visibility interventions can generate spillovers beyond the directly promoted items \citep{bairathi2025value}. However, whereas their mechanism operates through changes in users' perceptions of platform-wide quality, the patterns we document point to post-intervention recommender-policy updating: the system reallocates feed composition in a way that tracks the semantic structure of pre-existing user behavior.

This pattern is consistent with recommender-system learning. The intervention does not merely add a small number of reminder videos to the feed. Instead, the post-campaign feed shifts toward a broader set of entertainment-oriented attributes, as would occur if the system updated its representation of user preferences and reoptimized content allocation.

\subsection{Persistent Engagement Gains and Match Quality}
\label{subsec:engagement}

Implication 2 (Proposition~\ref{prop:2}) predicts that an intervention can raise engagement during the campaign if it moves exposure closer to latent preference, and can continue to raise engagement after the campaign if the intervention-period response triggers a policy update. The second prediction is central to Implication~3: post-campaign persistence should not depend on continued exposure to the reminder videos, because the score boost is no longer active. We first document this persistence pattern in the data, and then examine whether it could be generated by an alternative demand-side mechanism, in particular implicit advertising, rather than by recommender-system learning.

Section~\ref{sec:results} shows that the intervention's effect on total app usage and night usage persists for several weeks after the campaign ends. The ITT effect on night usage remains statistically significant at $p < 0.01$ through all eight post-intervention weeks, with the relative effect declining from 13.82\% in Week~0 to 3.60\% in Week~8 (Table~\ref{tab:longterm_nightusage}). For compliers, the estimated effect on night usage remains roughly 22--43\% higher than the control mean several weeks after the campaign. In the language of the framework, the system moves from an incumbent policy inside the inaction zone to a post-campaign policy informed by the response generated during the intervention period. Persistent post-intervention effects have also been documented in other platform settings, although prior mechanisms primarily emphasize user incentives or engagement states rather than recommender-system adaptation \citep{lu2023immediate}.

This persistence helps distinguish the recommender-learning mechanism from simpler user-side explanations. A \textit{persuasion} explanation, in which reminder videos convince users to value late-night entertainment more, would tend to predict effects concentrated during the intervention and declining as the message fades. A \textit{novelty} explanation predicts engagement concentrated on the reminder videos themselves. A \textit{salience} explanation predicts an immediate boost that disappears once the intervention is removed, with little reason for systematic changes in unrelated categories. The data are not well described by these patterns: effects persist after the boost ends, spill over systematically across 38 content categories. These patterns point toward system updating, while still leaving a more sophisticated demand-side alternative to consider: implicit advertising.

Specifically, the celebrity sleep-reminder videos may function as implicit \textit{advertising} for celebrity content and its adjacent categories. The advertising literature has documented that ads can generate cross-category spillovers by reminding consumers of related products \citep{sahni2016advertising} and that advertising effects can persist well beyond the campaign period \citep{shapiro2021tv}. An advertising channel could also rationalize an inverse-U pattern in heterogeneous effects if ads work less effectively on users who already strongly like or dislike the focal content. We therefore conduct additional analyses to distinguish the RS learning mechanism from an advertising channel.

To separate supply-side from demand-side responses, we decompose the treatment effect on category-level consumption into an extensive margin (what the RS serves) and an intensive margin (how users respond conditional on what they are served):
\[
\text{Played} \;=\; \underbrace{\text{Recommended}}_{\text{Extensive margin}} \;\times\; \underbrace{\text{Conversion Rate}}_{\text{Intensive margin}},
\]
where $\text{Conversion Rate} = \text{Played} / \text{Recommended}$. Our earlier results (Tables~\ref{tab:category_ratio}--\ref{tab:category_count}) already establish that the extensive margin shifts persistently in favor of celebrity-adjacent categories, providing direct evidence of policy updating. The question is whether the intensive margin also changes, which would indicate a demand-side shift consistent with advertising.

Table~\ref{tab:category_conversion} (Appendix) reports the treatment effect on conversion rates by category. Almost all categories exhibit a statistically significant positive effect, both during the experimental week and in Weeks 1--8. The intervention thus also improved the intensive margin: users became more likely to watch what was recommended to them, making a pure supply-side story incomplete. The critical question is whether this improvement follows the co-occurrence structure tied to celebrity content, as advertising would predict, or is diffuse across categories.

\begin{figure}[t!]
    \centering
    \includegraphics[width=0.7\textwidth]{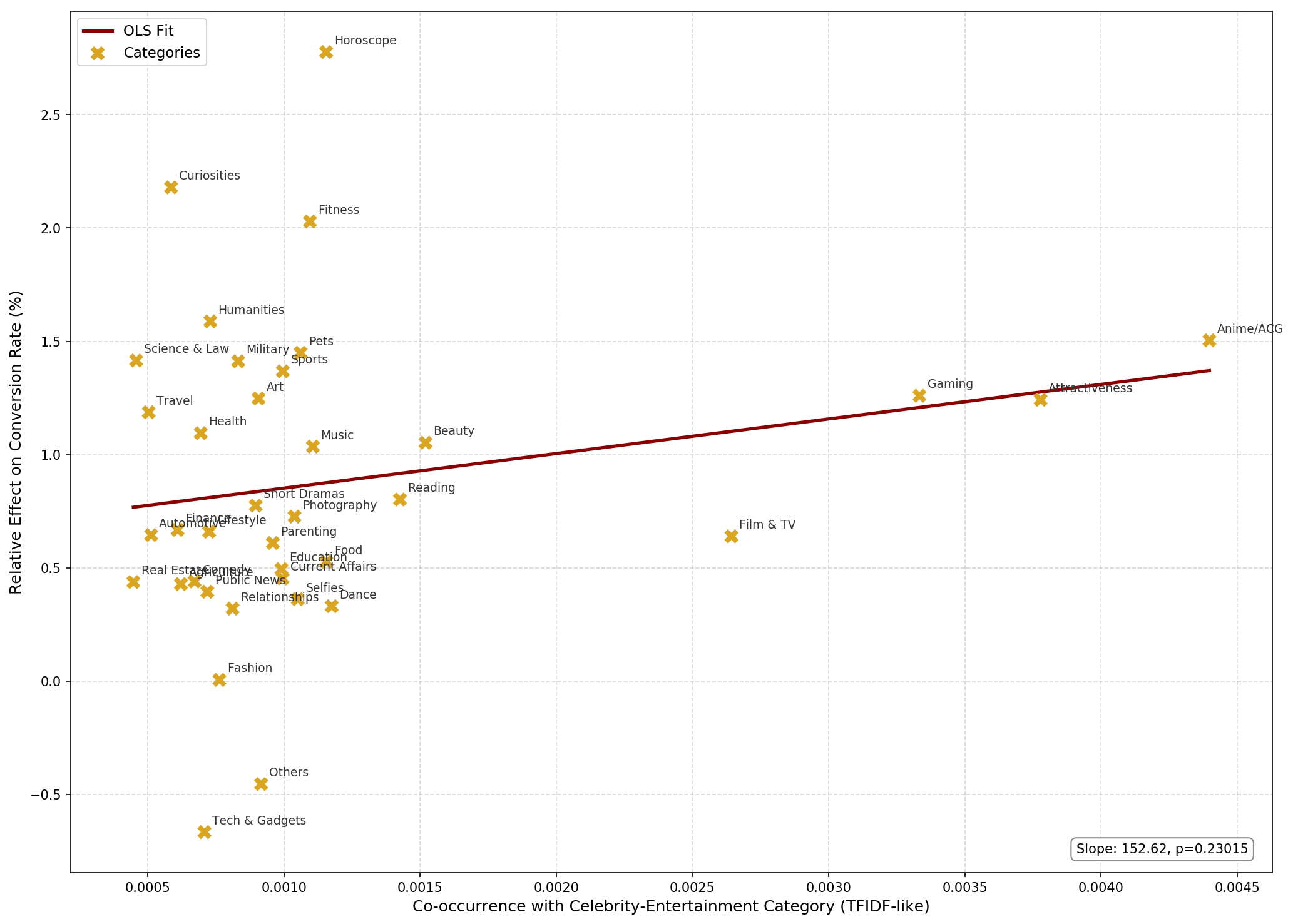}
    \caption{Conversion Rate Treatment Effect vs.\ Content Co-occurrence}
    \label{fig:cooccurrence_conversion}
    \begin{minipage}{0.9\textwidth}
        \scriptsize \textit{Notes:} Each point represents a content category. The y-axis shows the relative treatment effect on the conversion rate (played/recommended). The x-axis is a TF-IDF-like measure of the category's co-occurrence with the Celebrity-Entertainment category in user feeds pre-experiment. The flat slope (152.62, $p = 0.23$) indicates that conversion rate improvements do not follow the co-occurrence structure, in contrast to the significant relationship for recommended ratios (Figure~\ref{fig:cooccurrence_rec_ratio}; slope $= 797.20$, $p < 0.001$).
    \end{minipage}
\end{figure}
 
Figure~\ref{fig:cooccurrence_conversion} plots each category's conversion rate treatment effect against its co-occurrence score with the Celebrity-Entertainment category. In contrast to the highly significant relationship for recommended ratios (Figure~\ref{fig:cooccurrence_rec_ratio}; slope = 797.20, $p < 0.001$), the relationship for conversion rates is flat and insignificant (slope = 152.62, $p = 0.23$). Recommendation shares move in proportion to semantic relatedness to celebrity content, consistent with RS policy updating, while conversion-rate improvements are more diffuse, which weighs against category-specific advertising as the main explanation.

This test weighs against category-specific advertising, but a subtler version remains: the celebrity videos could strengthen users' overall connection to the platform, raising willingness to engage regardless of category. We cannot fully exclude this possibility, but several pieces of evidence limit its plausibility as the primary explanation. First, the treatment effect on live-stream viewing, which operates outside the short-video recommender system, is statistically insignificant throughout the study period (Section~\ref{sec:results}, Table~\ref{tab:shortterm_effects}). This ineffectiveness of the intervention on live-streaming is inconsistent with users developing a stronger \emph{overall} connection with the platform. Second, the ranking boost altered approximately 0.25\% of daily recommended content for treated users. Although direct comparisons of intervention dosage across settings are imperfect, advertising treatments shown in the literature to generate cross-category spillovers and persistent effects typically involve substantially more salient or frequent exposure than the ``0.25\%'' perturbation here \citep{sahni2016advertising, shapiro2021tv}.

In short, a more consistent explanation is better \textit{within-category} targeting: as the intervention generates new engagement data, the recommender updates not only cross-category allocation but also the matching of users to specific videos within categories. This channel explains why conversion rates can improve broadly while recommendation shares move in a structured way toward celebrity-adjacent categories. The evidence does not prove that advertising effects are absent, but it indicates that advertising alone is unlikely to account for the full pattern.

\subsection{Finite Attention and Reallocation Across Content Categories }
\label{subsec:finite_attention}

Because user attention and feed slots are finite, Implication 4 (Proposition~\ref{prop:3}) implies that policy updating can change category shares even when aggregate engagement increases. Empirically, this distinction between an engagement effect, which raises total usage, and a policy effect, which changes content allocation, is central to interpreting the observed spillovers. The logic that finite attention forces reallocation across categories connects to recent work showing that within-category satiation and cross-category spillovers shape user response in platform allocation systems \citep{lu2025within}.

Table~\ref{tab:category_count} reports treatment effects on the absolute number of recommended and played videos by category. Most categories experience increases in absolute counts, reflecting the overall rise in engagement. However, the magnitude of these increases varies substantially. Categories closely related to celebrity entertainment exhibit disproportionately large gains, while several unrelated categories experience smaller gains or declines in recommendation shares and, in some cases, play counts.

This pattern is consistent with finite-attention reallocation. As the recommender system updates its representation of what users value, it reoptimizes the feed by shifting category shares toward better-matched content. Even when the total amount of engagement grows, some categories can lose relative visibility because the policy change reallocates limited feed positions and user attention.

\begin{table}[t!]
    \centering
    \scriptsize
    \caption{Relative Effect Sizes of Treatment on Recommended and Played Counts (\%)}
    \label{tab:category_count}
    \begin{threeparttable}
    \begin{tabular}{l | c c c | c c c}
        \toprule
        \textbf{Category} 
        & \multicolumn{3}{c|}{\textbf{Recommended Count}} 
        & \multicolumn{3}{c}{\textbf{Played Count}} \\
        \cmidrule(lr){2-4} \cmidrule(lr){5-7}
        & \textbf{Baseline} 
        & \multicolumn{2}{c|}{\textbf{Relative Effect (\%)}} 
        & \textbf{Baseline} 
        & \multicolumn{2}{c}{\textbf{Relative Effect (\%)}} \\
        \cmidrule(lr){3-4} \cmidrule(lr){6-7}
        & 
        & \textbf{Week 0} 
        & \textbf{Week 1--8} 
        & 
        & \textbf{Week 0} 
        & \textbf{Week 1--8} \\
        \midrule
        Games & 17.47 & 7.28*** & 6.83*** & 6.51 & 9.96*** & 8.81*** \\
        Astrology \& Fortune-telling & 0.82 & 6.68*** & 3.61*** & 0.27 & 6.09*** & 3.27*** \\
        Appearance / Looks & 12.03 & 5.54*** & 4.95*** & 1.72 & 7.73*** & 6.44*** \\
        Photography & 2.90 & 5.12*** & 3.41*** & 0.69 & 5.10*** & 3.27*** \\
        Celebrity Entertainment & 8.42 & 4.86*** & 2.14*** & 2.95 & 5.68*** & 3.29*** \\
        Emotions / Relationships & 14.98 & 4.74*** & 3.87*** & 4.53 & 4.25*** & 3.39*** \\
        Tech \& Digital Products & 4.20 & 4.65*** & 3.70*** & 1.35 & 1.65*** & 1.19*** \\
        Sports & 7.30 & 4.38*** & 3.61*** & 2.83 & 5.63*** & 4.27*** \\
        Anime / ACG & 15.77 & 4.37*** & 4.22*** & 5.35 & 4.78*** & 4.45*** \\
        Film \& TV Shows & 32.80 & 4.15*** & 1.65*** & 13.33 & 4.89*** & 2.17*** \\
        Selfies & 33.88 & 3.30*** & 2.74*** & 5.09 & 4.70*** & 3.65*** \\
        Lifestyle & 10.48 & 2.79*** & 1.24*** & 3.16 & 3.14*** & 1.40*** \\
        Comedy & 18.04 & 2.75*** & 2.13*** & 8.28 & 3.47*** & 2.51*** \\
        Oddities \& Wonders & 1.93 & 2.72*** & 1.40*** & 0.84 & 4.08*** & 1.89*** \\
        Beauty / Makeup & 7.79 & 2.54*** & 1.97*** & 1.81 & 3.97*** & 3.56*** \\
        Others & 26.44 & 2.27*** & 1.57*** & 4.27 & 3.05*** & 2.43*** \\
        Education & 2.68 & 1.88** & 0.54** & 0.82 & 1.71*** & 0.36 \\
        Short Dramas & 11.74 & 1.79*** & 1.84*** & 4.94 & 2.07*** & 1.74*** \\
        Fashion & 20.60 & 1.65*** & -0.12 & 3.77 & 1.58*** & 0.49*** \\
        Finance \& Business & 1.98 & 1.53*** & 0.30 & 0.71 & 0.55 & -0.30 \\
        Humanities & 2.31 & 1.51*** & -0.60*** & 0.92 & 1.33** & 0.04 \\
        Current Affairs \& News & 1.42 & 1.42** & 0.94*** & 0.72 & -0.04 & 0.81*** \\
        Science \& Law & 2.19 & 1.32*** & 0.32* & 0.90 & 1.87*** & 0.94*** \\
        Civic News & 16.56 & 1.19*** & 0.52*** & 7.94 & 1.79*** & 1.26*** \\
        Reading / Books & 3.45 & 1.19** & -0.80*** & 1.29 & 0.87 & -0.64*** \\
        Music & 13.16 & 1.15*** & 0.81*** & 3.54 & 0.69 & 0.13 \\
        Travel & 4.11 & 0.68* & -0.04 & 1.01 & 2.49*** & 1.40*** \\
        Dance & 8.61 & 0.64 & 1.11*** & 1.87 & 1.36*** & 0.90*** \\
        Fitness & 1.40 & 0.59 & 1.36*** & 0.37 & 0.81 & 0.65* \\
        Military & 2.52 & 0.57 & 0.57** & 1.25 & 0.43 & 1.20*** \\
        Pets & 4.42 & 0.47 & 0.46* & 1.60 & 2.28*** & 1.55*** \\
        Health & 5.97 & 0.17 & -0.22 & 1.88 & 0.80* & 0.20 \\
        Parenting \& Family & 9.31 & 0.09 & 0.49*** & 3.01 & -0.18 & 0.16 \\
        Real Estate \& Home & 7.55 & 0.02 & -1.23*** & 2.63 & 0.82** & -0.47*** \\
        Food & 20.51 & -0.25 & 0.27* & 5.86 & 1.41*** & 1.00*** \\
        Art & 6.32 & -0.62 & -0.28 & 2.13 & 1.62*** & 0.99*** \\
        Automobiles & 10.07 & -1.50** & -2.45*** & 2.79 & 1.84*** & -0.09 \\
        Rural / Agriculture & 11.66 & -1.66*** & -1.20*** & 3.33 & -0.08 & -0.44*** \\
        \bottomrule
    \end{tabular}
    \begin{tablenotes}
    \scriptsize
    \item Note: This table shows the treatment's effects on the absolute number of videos from different content categories. The analysis is at the user-day-category level. ``Baseline" columns show the average number of videos from each category for the control group. ``Relative Effect (\%)" columns report the ITT effect as a percentage, calculated from our main OLS specification as (coefficient on Treatment / baseline) * 100. The effects are shown separately for the experimental week (Week 0) and the following eight weeks (Week 1-8). $^{*}$p$<$0.1; $^{**}$p$<$0.05; $^{***}$p$<$0.01.
    \end{tablenotes}
    \end{threeparttable}
\end{table}

\subsection{Heterogeneous Effects and Embedding Bias}
\label{subsec:heterogeneity}

Implication 5 (Proposition~\ref{prop:4}) suggests that the magnitude of updating should vary with the initial misalignment between the platform's learned representation and the user's latent preference, as well as with the informativeness of the intervention-generated signal. We cannot observe this misalignment directly. We therefore proxy for the scope for updating using the platform's pre-experiment recommendation behavior. Specifically, we group users into quintiles based on their pre-experiment celebrity-content recommendation exposure, calculated as the share of celebrity videos in their recommended feed in the week before the experiment.

We expect the largest treatment effects among users with moderate pre-experiment celebrity exposure. At the top of the distribution (Q1), the platform is already serving substantial celebrity-related content, so its belief about the user's preference for the category is likely already well aligned, and the intervention provides confirmatory rather than corrective information. In the language of the framework, $|\theta-m_0|$ is small for these users, so the update margin in Implication 5 (Proposition~\ref{prop:4}) is small. At the bottom of the distribution (Q4--Q5), the platform's belief is also well resolved, either because the prior is accurate (the user genuinely has low affinity for the focal category) or because it is held with high precision. In either case, the intervention generates limited new information and induces a small policy update. The moderate-exposure groups (Q2--Q3) are most likely to contain users for whom the system has some evidence of interest but has not fully resolved the preference, leaving meaningful scope for an intervention-induced update.

To test this prediction, we estimate the following fully interacted model to examine heterogeneous treatment effects (HTE):
\begin{equation}
\text{Ratio}_{it} = \beta_0 + \sum_{q=2}^{5} \delta_q Q_{iq} + \beta_1 \text{Treat}_i + \sum_{q=2}^{5} \gamma_q (Q_{iq} \times \text{Treat}_i) + \epsilon_{it}
\end{equation}
where $\text{Ratio}_{it}$ is the share of celebrity content in user $i$'s recommended feed on day $t$ during the experiment week. $Q_{iq}$ is an indicator for user $i$ belonging to pre-treatment celebrity recommendation exposure quintile $q, q\in\{1,2,...,5\}$, where quintile 1 received the highest share of celebrity content recommendations and quintile 5 received the lowest share.

Table \ref{tab:hte_celebrity_exposure_ratio} presents heterogeneous treatment effects on the share of celebrity content in users' recommendation feeds across these quintiles. The results show that the treatment effect is not uniform. The relative effect on recommended celebrity content ratio is most pronounced and positive for the second and third quintiles (users with a moderate level of initial recommendation exposure), with relative effects of 1.13\% and 1.71\%, respectively.

The resulting non-monotonic pattern is consistent with Implication 5 (Proposition~\ref{prop:4}) under the mapping just described. Q1 users show a near-zero relative effect ($-$0.05\%), consistent with the platform already serving substantial celebrity content to this group. Q2--Q3 users show the largest positive relative effects (1.13\% and 1.71\%), suggesting that the intervention generated the most useful information for this moderate-exposure segment. Q4--Q5 users show smaller positive effects (0.62\%), consistent with limited room for updating at the bottom of the exposure distribution. The concentration of effects in the middle of the distribution is difficult to reconcile with a homogeneous salience or reminder effect, and is more consistent with a mechanism in which the value of the intervention depends on the system's pre-existing representation.

\begin{table}[t!]
  \centering
\scriptsize
  \begin{threeparttable}
  \caption{HTE on Celebrity Content Recommendations by Pre-Campaign Exposure}
  \label{tab:hte_celebrity_exposure_ratio}
    \begin{tabular}{l c c}
    \toprule
    & \multicolumn{2}{c}{\textbf{Celebrity Content Ratio}} \\
    \cmidrule(lr){2-3}
    & \textbf{All Videos} & \textbf{Top 20 Videos} \\
    \midrule
    Constant & 0.0592$^{***}$ & 0.0583$^{***}$ \\
           & (0.0001)       & (0.0001) \\
    2nd Quintile & -0.0354$^{***}$ & -0.0356$^{***}$ \\
           & (0.0001)       & (0.0001) \\
    3rd Quintile & -0.0434$^{***}$ & -0.0432$^{***}$ \\
           & (0.0001)       & (0.0001) \\
    4th Quintile & -0.0480$^{***}$ & -0.0476$^{***}$ \\
           & (0.0001)       & (0.0001) \\
    5th Quintile & -0.0518$^{***}$ & -0.0512$^{***}$ \\
           & (0.0001)       & (0.0001) \\
    Treat & -0.0000 & -0.0001 \\
           & (0.0001) & (0.0001) \\
    Treat $\times$ 2nd Quintile & 0.0003 & 0.0002 \\
           & (0.0002) & (0.0002) \\
    Treat $\times$ 3rd Quintile & 0.0003 & 0.0002 \\
           & (0.0002) & (0.0002) \\
    Treat $\times$ 4th Quintile & 0.0001 & 0.0001 \\
           & (0.0002) & (0.0002) \\
    Treat $\times$ 5th Quintile & 0.0001 & 0.0001 \\
           & (0.0002) & (0.0002) \\
    \midrule
    \textit{Implied Relative Effects (\%)} \\
    Relative Effect (1st Quintile, highest exposure) & -0.05\% & -0.12\% \\
    Relative Effect (2nd Quintile) & 1.13\% & 0.57\% \\
    Relative Effect (3rd Quintile) & 1.71\% & 0.85\% \\
    Relative Effect (4th Quintile) & 0.62\% & -0.19\% \\
    Relative Effect (5th Quintile, lowest exposure) & 0.62\% & -0.07\% \\
    \midrule
    Observations & 1404403 & 1404403\\
    \bottomrule
    \end{tabular}
    \begin{tablenotes}
    \scriptsize
    \item Note: This table presents heterogeneous treatment effects on the share of celebrity entertainment videos in users' recommended feeds. Users are grouped into quintiles based on their pre-campaign celebrity recommendation exposure, measured as the share of celebrity videos in their recommended feed during the week before the experiment. The first quintile has the highest pre-campaign exposure and the fifth quintile has the lowest. The analysis is at the user-day level for the experimental week (April 29--May 6, 2022). The top panel reports coefficients from a fully interacted OLS model. The bottom panel calculates the implied relative treatment effect for each quintile. For Quintile 1, the effect is (Treat / Constant) * 100. For Quintiles 2--5, the effect is ((Treat + Treat $\times$ Quintile) / (Constant + Quintile)) * 100. Robust standard errors are reported in parentheses. $^{*}$p$<$0.1; $^{**}$p$<$0.05; $^{***}$p$<$0.01.
    \end{tablenotes}
  \end{threeparttable}
\end{table}

In sum, the empirical results are consistent with the forced-exploration mechanism. The sleep-reminder intervention was designed to curb late-night engagement, but it also generated behavioral responses to content that the baseline policy had not served as often. The recommender system appears to have used these responses to adjust subsequent content allocation, improving match quality and increasing engagement after the campaign.

Taken together, these analyses make purely user-side explanations less plausible. Persuasion, novelty, and salience predict short-lived and localized effects, whereas the data show persistent, cross-category, and heterogeneous patterns. A more sophisticated advertising alternative can account for some persistence and spillovers, but the co-occurrence tests show that recommendation shares, not conversion-rate improvements, follow the celebrity-related content structure. Habit formation would predict repetition of pre-existing viewing patterns rather than structured reshaping of content exposure across 38 categories. The forced-exploration interpretation provides the most parsimonious account of the joint pattern observed in the data: a temporary intervention creates a training episode, and the post-campaign empirical signatures---shifts in feed allocation, improved within-category matching, and reallocation of finite attention across the content ecosystem---are consistent with the system updating its representation in response.

\section{Spillovers and SUTVA}\label{sec:sutva}

A foundational assumption in randomized experiments is the Stable Unit Treatment Value Assumption (SUTVA), which requires that one unit's treatment assignment does not affect another unit's outcomes. However, in a platform setting in which the recommender system pools data across users, the system itself is a plausible interference channel, and we discuss the primary mechanisms below.

The most direct concern is RS-mediated model updating. The platform's recommender trains on consumption feedback pooled across users. To the extent that signals generated by treated users are incorporated into shared model components such as item embeddings or global scoring parameters, control users' recommendations may shift in the same direction predicted by the intervention. A related channel operates through shared content pools: if the intervention raises engagement with particular videos among treated users, those videos may accumulate higher organic scores and thereby rank higher for control users as well. Both channels would attenuate the estimated treatment-control difference, since the control group partially inherits any belief correction or popularity gain induced by the intervention.

Several features of the experimental setting limit the scope of these spillovers. First, only 10\% of users were assigned to the treatment group, so treatment-induced signals constitute a small share of the total training data feeding the platform's model updates. Second, the intervention itself altered approximately 0.25\% of daily recommended content for treated users, a small perturbation relative to the volume of organic behavior the platform observes each day. Third, the platform updates its models in batch cycles rather than in real time, introducing a temporal buffer between treated users' behavioral responses and any downstream effect on recommendation policy. Fourth, the user embeddings that drive personalized ranking are primarily shaped by each user's own consumption history, which limits the degree to which one user's intervention-induced behavior can directly alter another user's feed composition.

These design features reduce the scope for interference but do not eliminate it. To the extent that SUTVA is violated through either the user-side learning channel or shared content pools, the bias points in a known direction: control-group recommendations and engagement also shift toward the post-intervention policy, attenuating the estimated treatment-control difference. The estimates reported in Sections~\ref{sec:results} and \ref{sec:mechanism} should therefore be interpreted as conservative lower bounds on the magnitude of the intervention's effect on the recommender system's adaptation.

\section{Conclusion}\label{sec:conclusion}

This study investigates the unintended consequences of a content intervention implemented within a complex recommender system. We document and explain a paradoxical outcome: a ``sleep reminder'' campaign, designed to reduce late-night usage on a major short-video platform, instead caused a sustained increase in both late-night and overall user engagement. Notably, this effect arises despite the fact that the intervention altered the ranking of only about 0.25 percent of daily recommended videos for treated users, yet the effects on subsequent engagement are substantial. We propose and empirically validate a mechanism, in which a temporary intervention changes exposure, generates new user responses, and gives the recommender system evidence that can justify a post-campaign policy update. Under this mechanism, the system may otherwise remain in an inaction zone because routine feedback is insufficient to support an update. The intervention can move the system out of that region and lead to a recommendation policy with improved content-user matching and higher engagement.

Our findings make several important contributions. Most centrally, we document evidence consistent with an ``algorithmic nudge reversal,'' a phenomenon in which the adaptive learning dynamics of a recommender system overturn the intended direction of a digital well-being intervention. This adds a dynamic and system-level perspective to the literature on platform interventions, algorithmic feedback loops, and bias correction, highlighting how algorithmic feedback loops can trap systems in local optima by censoring the data necessary for learning. Rather than serving as a passive conduit for policy changes, the recommender system behaves in a manner consistent with actively learning from the intervention, reshaping future content allocation, and ultimately driving long-run outcomes.

The results also yield important managerial implications. First, the evaluation of platform interventions requires a system-level perspective and a sufficiently long time horizon. Short-time-window evaluations of A/B tests, even when methodologically clean, may miss slower but more consequential effects that emerge through algorithmic adaptation. Managers must therefore account not only for how users respond to an intervention, but also for how the system itself learns from it. Second, our findings suggest that content interventions can be reconceptualized as strategic tools for system calibration. Rather than treating them solely as behavioral nudges, platforms could design them as targeted exploration campaigns to resolve identification problems, probe latent user preferences, reveal structural biases in their models, and identify previously under-served areas of demand. We note that this second implication operates in tension with the welfare-oriented motivation of the original campaign: the same forced-exploration logic that makes interventions useful as calibration tools also makes them potential drivers of higher engagement, complicating their use as instruments for user well-being.

Finally, this study highlights an inherent tension in the pursuit of digital well-being on algorithmic platforms. Even interventions explicitly intended to promote healthier usage patterns can activate learning mechanisms that improve recommendation accuracy and make the platform more engaging. This paradox complicates corporate social responsibility initiatives and calls for more nuanced intervention designs that explicitly anticipate and address algorithmic feedback effects.

Our research is subject to several limitations that point to promising directions for future work. First, we study a specific type of intervention, namely the boosting of celebrity-recorded reminder content, on a short-video platform. Future research should examine whether the algorithmic nudge reversal generalizes to other intervention types, such as interface changes or the de-boosting of harmful content, and to other digital contexts, including e-commerce, news aggregation, and online dating. Second, while the formal model in Appendix A captures the core mechanism in a tractable way, it necessarily abstracts from the full complexity of industrial recommender systems. Future work could incorporate richer features such as network effects, multi-dimensional preference embeddings, and more detailed representations of model training pipelines. Finally, although we focus on behavioral engagement outcomes, the broader implications for user welfare remain an open question. Future studies could combine experimental methods with survey data or well-being measures to directly assess how algorithm-mediated changes in exposure shape users’ subjective experiences and long-term outcomes. Such research would be critical for guiding platforms as they navigate the trade-offs between optimizing engagement and genuinely promoting user well-being.

\bibliographystyle{informs2014} 
\bibliography{ref} 








\section*{Appendix A: Formal Model and Proofs}

This appendix formalizes the mechanism described in Section~\ref{sec:model}. The model follows a representative user segment through three periods that correspond to the pre-campaign, campaign, and post-campaign windows in the empirical setting. The goal is to characterize how a temporary intervention can generate new feedback and affect the post-campaign recommendation policy after the intervention itself has ended.

\subsection*{A.1 Preferences and Recommendation Policy}

A representative user segment has a latent preference $\theta\in[0,1]$ for a focal content category. In the empirical setting, the focal category corresponds to celebrity-related reminder content and nearby content attributes. The platform chooses a scalar recommendation policy $\lambda\in[0,1]$, which denotes the effective intensity or share of focal-category content in the feed. The experimental score boost maps into this policy by increasing the probability that focal content clears the ranking threshold.

The match quality between the user segment and the recommendation policy is
\begin{equation}
A(\theta,\lambda)=1-(\theta-\lambda)^2.
\label{eq:app_match_quality}
\end{equation}
Expected consumption is
\begin{equation}
G(\lambda;\theta)=h\!\left(1-(\theta-\lambda)^2\right),
\label{eq:app_segment_engagement}
\end{equation}
where $h$ is twice continuously differentiable, $h(A)\geq0$, $h'(A)>0$, and $h''(A)\leq0$ on $[0,1]$. These assumptions impose positive and weakly diminishing returns to match quality. One microfoundation is a user problem $\max_{x\geq0} Ax-c(x)$ with $c''(x)>0$ and an interior solution, in which case $x=h(A)$.

The expected-consumption function is strictly concave in $\lambda$. To see this, note that
\begin{equation}
\frac{\partial^2 G(\lambda;\theta)}{\partial \lambda^2}
=
4h''(A)(\theta-\lambda)^2-2h'(A)<0.
\label{eq:app_G_second}
\end{equation}
The first term is weakly nonpositive and the second term is strictly negative. Hence $G(\lambda;\theta)$ is uniquely maximized at $\lambda=\theta$.

\subsection*{A.2 Timing, Feedback, and Posterior Beliefs}

There are three periods. Period 0 is the pre-intervention baseline. Period 1 is the intervention period. Period 2 is the post-intervention period. Let $B$ denote the no-intervention regime and $I$ denote the intervention regime.

In period 0, the platform does not observe $\theta$ directly. Its incumbent representation of the segment is summarized by a prior mean $m_0\in[0,1]$ and precision $\tau_m>0$. The baseline recommendation policy is
\begin{equation}
\lambda_0=m_0.
\label{eq:app_baseline_policy}
\end{equation}
Embedding bias is present when $m_0\neq\theta$.

In period 1, the no-intervention regime continues to use the incumbent policy. The intervention regime applies a temporary perturbation to the focal category. The two period-1 policies are
\begin{equation}
\lambda_1^B=m_0,
\qquad
\lambda_1^I=m_0+a,
\qquad
0<a\leq 1-m_0.
\label{eq:app_period1_policies}
\end{equation}
The constraint on $a$ ensures that the intervention policy remains in $[0,1]$.

Period-1 behavior generates realized feedback
\begin{equation}
Y_1^r=G(\lambda_1^r;\theta)+\eta_1^r,
\qquad
\mathbb{E}[\eta_1^r\mid\theta]=0,
\qquad
r\in\{B,I\}.
\label{eq:app_noisy_consumption}
\end{equation}
The platform uses this feedback, together with the associated interaction logs, to form a reduced-form preference signal
\begin{equation}
s_1^r=\theta+\varepsilon_1^r,
\qquad
\mathbb{E}[\varepsilon_1^r\mid\theta]=0,
\qquad
\mathrm{Var}(\varepsilon_1^r\mid\theta)=(\tau_1^r)^{-1}
\quad\text{when }\tau_1^r>0.
\label{eq:app_signal}
\end{equation}
The precision $\tau_1^r$ summarizes how informative the period's data are about the latent preference. This reduced-form signal avoids modeling the full feature-extraction and retraining pipeline while preserving the information channel that matters for the update decision.

Routine feedback has precision $\tau_1^B\geq0$, with the corner case $\tau_1^B=0$ corresponding to fully censored routine feedback that carries no preference information. The intervention is more informative because it exposes the segment to focal content that the baseline policy would have served less often. We assume
\begin{equation}
\tau_1^I(a)>\tau_1^B,
\qquad
\frac{d\tau_1^I(a)}{da}>0
\quad\text{for }a\in(0,1-m_0].
\label{eq:app_precision_order}
\end{equation}
Given a realized period-1 signal, the posterior mean is
\begin{equation}
\widetilde m_2^r(s_1^r)=(1-\omega^r)m_0+\omega^r s_1^r,
\qquad
\omega^r=\frac{\tau_1^r}{\tau_m+\tau_1^r},
\qquad
r\in\{B,I\}.
\label{eq:app_random_posterior}
\end{equation}
Thus $\widetilde m_2^r(s_1^r)$ is a random variable before the period-1 signal is realized.

The propositions characterize the mean-signal path. Along this path, the realized signal equals its conditional mean, $s_1^r=\mathbb{E}[s_1^r\mid\theta]=\theta$. Equivalently, the analysis describes the deterministic limit in which the period-1 signal aggregates many conditionally independent observations. The posterior means used in the deterministic comparisons are therefore
\begin{equation}
m_2^B=(1-\omega^B)m_0+\omega^B\theta,
\qquad
\omega^B=\frac{\tau_1^B}{\tau_m+\tau_1^B},
\label{eq:app_routine_update}
\end{equation}
and
\begin{equation}
m_2^I=(1-\omega^I(a))m_0+\omega^I(a)\theta,
\qquad
\omega^I(a)=\frac{\tau_1^I(a)}{\tau_m+\tau_1^I(a)}.
\label{eq:app_intervention_update}
\end{equation}
Because $\tau_m>0$ and $\tau_1^I(a)>0$, $\omega^I(a)\in(0,1)$. Because $\tau_1^I(a)>\tau_1^B$, $\omega^I(a)>\omega^B$.

In period 2, the intervention boost is no longer applied. The platform decides whether to keep the incumbent policy or update the policy using the period-1 signal.

\subsection*{A.3 Updating and the Inaction Zone}

Policy changes are costly. The cost $K>0$ represents retraining costs, engineering frictions, governance thresholds, or other reasons why platforms do not revise recommendation policies after every small signal. We use a myopic certainty-equivalent update rule as a tractable representation of these update frictions; we do not interpret $K$ as a literal engineering cost or the resulting threshold $\Delta_K$ derived below as a directly observable parameter of the platform's retraining pipeline. Given a candidate posterior mean $m$, the platform treats $m$ as a point estimate of the latent preference and compares the payoff from staying with the incumbent policy $m_0$ to the payoff from updating to the policy that is myopically optimal under that point estimate:
\begin{equation}
\text{update iff } G(m;m)-G(m_0;m)>K.
\label{eq:app_update_decision}
\end{equation}
Since $G(m;m)=h(1)$ and $G(m_0;m)=h(1-(m-m_0)^2)$, the gross gain from updating is
\begin{equation}
h(1)-h\!\left(1-(m-m_0)^2\right).
\label{eq:app_update_gain}
\end{equation}
Assume $K\in(0,h(1)-h(0))$ and ties are broken in favor of the incumbent policy. Since $h$ is strictly increasing, define
\begin{equation}
\Delta_K
\equiv
\sqrt{1-h^{-1}(h(1)-K)}.
\label{eq:app_inaction_threshold}
\end{equation}
The update rule is equivalent to
\begin{equation}
|m-m_0|>\Delta_K.
\label{eq:app_inaction_rule}
\end{equation}
Thus, the interval of posterior means satisfying $|m-m_0|\leq\Delta_K$ is an inaction zone around the incumbent policy.

For each regime $r\in\{B,I\}$, the period-2 policy along the mean-signal path is
\begin{equation}
\lambda_2^r=
\begin{cases}
 m_2^r, & \text{if } |m_2^r-m_0|>\Delta_K,\\
 m_0, & \text{if } |m_2^r-m_0|\leq\Delta_K.
\end{cases}
\label{eq:app_period2_policy}
\end{equation}

\begin{proposition}[Baseline Inaction and Intervention-Induced Updating]\label{prop:1}
Under routine feedback, the platform remains with the incumbent policy if and only if
\begin{equation}
\omega^B|\theta-m_0|\leq \Delta_K.
\label{eq:app_baseline_inaction_condition}
\end{equation}
Under intervention-period feedback, the period-2 policy updates if and only if
\begin{equation}
\omega^I(a)|\theta-m_0|> \Delta_K.
\label{eq:app_intervention_update_condition}
\end{equation}
If both inequalities hold, routine feedback leaves $\lambda_2^B=m_0$, while intervention-generated feedback induces $\lambda_2^I=m_2^I$. If $m_0<\theta$, then $m_0<m_2^I<\theta$.
\end{proposition}

\begin{proposition}[Immediate and Persistent Consumption Effects]\label{prop:2}
Suppose the platform initially underestimates focal-category preference, $m_0<\theta$, and the intervention raises focal-content intensity without overshooting:
\begin{equation}
0<a\leq \theta-m_0.
\label{eq:app_no_overshooting}
\end{equation}
Then period-1 consumption under the intervention is higher than period-1 consumption under the baseline policy:
\begin{equation}
G(\lambda_1^I;\theta)>G(\lambda_1^B;\theta)=G(m_0;\theta).
\label{eq:app_immediate_consumption}
\end{equation}
If, in addition, the threshold-crossing conditions in Proposition~\ref{prop:1} hold, then post-intervention consumption in period 2 is higher under the intervention-induced policy:
\begin{equation}
G(\lambda_2^I;\theta)>G(\lambda_2^B;\theta).
\label{eq:app_persistent_consumption}
\end{equation}
\end{proposition}

\subsection*{A.4 Category-Level Accounting}

The scalar model above characterizes total expected consumption. The empirical analysis also examines how that consumption is distributed across focal and non-focal categories. To map the scalar policy into category-level accounting, we use a proportional allocation benchmark: when the feed contains share $\lambda$ of focal content and total expected consumption is $G(\lambda;\theta)$, focal-category consumption and non-focal consumption are
\begin{equation}
C(\lambda)=\lambda G(\lambda;\theta),
\qquad
NC(\lambda)=(1-\lambda)G(\lambda;\theta).
\label{eq:app_proportional_allocation}
\end{equation}
This benchmark treats $\lambda$ as both the focal share of the recommendation policy and the focal share used to decompose total consumption into category-level components.\footnote{We adopt proportional allocation for tractability. The empirical analysis in Section~\ref{subsec:engagement} examines conversion-rate heterogeneity directly and uses departures from uniform conversion to distinguish algorithmic updating from advertising effects on user demand. That analysis does not rely on the proportional-allocation benchmark used in Proposition~\ref{prop:3}.}

\begin{proposition}[Finite Attention and Content Reallocation]\label{prop:3}
Suppose $\theta>m_0$. Consider any post-update policy $\lambda_2\in[0,1]$ such that $\lambda_2>m_0$ and $G(\lambda_2;\theta)>G(m_0;\theta)$. Under proportional allocation, non-focal consumption decreases, $NC(\lambda_2)<NC(m_0)$, if and only if
\begin{equation}
(\lambda_2-m_0)G(m_0;\theta)
>
(1-\lambda_2)\bigl[G(\lambda_2;\theta)-G(m_0;\theta)\bigr].
\label{eq:app_reallocation_condition}
\end{equation}
\end{proposition}

\subsection*{A.5 Determinants of Post-Intervention Updating}

The preceding results show that a temporary intervention can have persistent consequences when the intervention-generated posterior crosses the update threshold. To characterize when this crossing is more likely, define the intervention update margin as
\begin{equation}
M(a,\theta,m_0,K)
=
\omega^I(a)|\theta-m_0|-\Delta_K.
\label{eq:app_update_margin}
\end{equation}

\begin{proposition}[Update Margin and Comparative Statics]\label{prop:4}
The intervention triggers a post-intervention update along the mean-signal path if and only if $M(a,\theta,m_0,K)>0$. Moreover, the update margin is strictly increasing in the magnitude of embedding bias $|\theta-m_0|$, weakly increasing in intervention-generated signal precision $\tau_1^I(a)$ (strictly so when $\theta\neq m_0$), and strictly decreasing in the update cost $K$. Over any interval with $a\in(0,1-m_0]$ and $d\tau_1^I(a)/da>0$, the update margin is weakly increasing in the intervention intensity $a$, and it is strictly increasing when $\theta\neq m_0$.
\end{proposition}

\subsection*{A.6 Proofs}

\subsection*{Proof of Proposition \ref{prop:1}}

By \eqref{eq:app_period2_policy}, routine feedback leaves the platform with the incumbent policy if and only if $|m_2^B-m_0|\leq\Delta_K$. Along the mean-signal path,
\begin{equation}
|m_2^B-m_0|
=
\left|(1-\omega^B)m_0+\omega^B\theta-m_0\right|
=
\omega^B|\theta-m_0|.
\label{eq:app_proof_baseline_move}
\end{equation}
This establishes \eqref{eq:app_baseline_inaction_condition}. The corresponding movement under intervention-period feedback is
\begin{equation}
|m_2^I-m_0|
=
\omega^I(a)|\theta-m_0|,
\label{eq:app_proof_intervention_move}
\end{equation}
so the platform updates if and only if \eqref{eq:app_intervention_update_condition} holds. If both conditions hold, \eqref{eq:app_period2_policy} implies $\lambda_2^B=m_0$ and $\lambda_2^I=m_2^I$. Finally, when $m_0<\theta$, $\tau_m>0$, and $\tau_1^I(a)>0$, we have $\omega^I(a)\in(0,1)$. Therefore
\begin{equation}
m_2^I=(1-\omega^I(a))m_0+\omega^I(a)\theta
\end{equation}
is a strict convex combination of $m_0$ and $\theta$. Hence $m_0<m_2^I<\theta$. \Halmos

\subsection*{Proof of Proposition \ref{prop:2}}

Because $G(\lambda;\theta)=h(1-(\theta-\lambda)^2)$ and $h$ is strictly increasing, $G(\lambda;\theta)$ increases whenever $|\theta-\lambda|$ decreases. If $m_0<\theta$ and $0<a\leq\theta-m_0$, then
\begin{equation}
|\theta-(m_0+a)|<|\theta-m_0|,
\end{equation}
with strict inequality because $a>0$. Hence
\begin{equation}
G(m_0+a;\theta)>G(m_0;\theta),
\end{equation}
which proves the immediate consumption effect. If the threshold-crossing conditions in Proposition~\ref{prop:1} hold, then $\lambda_2^B=m_0$ and $\lambda_2^I=m_2^I$. From Proposition~\ref{prop:1}, $m_0<m_2^I<\theta$, so
\begin{equation}
|\theta-m_2^I|<|\theta-m_0|.
\end{equation}
Strict monotonicity of $h$ then implies $G(m_2^I;\theta)>G(m_0;\theta)$, which establishes \eqref{eq:app_persistent_consumption}. \Halmos

\subsection*{Proof of Proposition \ref{prop:3}}

Non-focal consumption under policy $\lambda$ is $NC(\lambda)=(1-\lambda)G(\lambda;\theta)$. Therefore,
\begin{align}
NC(\lambda_2)-NC(m_0)
&=(1-\lambda_2)G(\lambda_2;\theta)-(1-m_0)G(m_0;\theta)\\
&=(1-\lambda_2)\bigl[G(\lambda_2;\theta)-G(m_0;\theta)\bigr]-(\lambda_2-m_0)G(m_0;\theta).
\label{eq:app_nc_decomposition}
\end{align}
Thus $NC(\lambda_2)<NC(m_0)$ if and only if
\begin{equation}
(1-\lambda_2)\bigl[G(\lambda_2;\theta)-G(m_0;\theta)\bigr]
<
(\lambda_2-m_0)G(m_0;\theta),
\end{equation}
which is equivalent to \eqref{eq:app_reallocation_condition}.

To see that the condition is not vacuous, consider the following example with
\(h(A)=A\), \(\theta=0.9\), \(m_0=0.5\), and \(\lambda_2=0.85\). Then
\(G(m_0;\theta)=1-(0.9-0.5)^2=0.84\), while
\(G(\lambda_2;\theta)=1-(0.9-0.85)^2=0.9975\). Total expected consumption
therefore increases. Under proportional allocation, however,
\(NC(m_0)=0.5\times0.84=0.42\), whereas
\(NC(\lambda_2)=0.15\times0.9975=0.149625\). Equivalently, the left-hand side
of \eqref{eq:app_reallocation_condition} equals
\((\lambda_2-m_0)G(m_0;\theta)=0.294\), while the right-hand side equals
\((1-\lambda_2)\{G(\lambda_2;\theta)-G(m_0;\theta)\}=0.023625\). Thus
\eqref{eq:app_reallocation_condition} holds strictly.
\Halmos

\subsection*{Proof of Proposition \ref{prop:4}}

By Proposition~\ref{prop:1}, intervention-generated feedback induces a period-2 update along the mean-signal path if and only if $\omega^I(a)|\theta-m_0|>\Delta_K$, which is equivalent to $M(a,\theta,m_0,K)>0$. The update margin is linear in $|\theta-m_0|$ with coefficient $\omega^I(a)>0$, so it is strictly increasing in the magnitude of embedding bias. Since
\begin{equation}
\omega^I(a)=\frac{\tau_1^I(a)}{\tau_m+\tau_1^I(a)},
\end{equation}
we have
\begin{equation}
\frac{\partial M}{\partial \tau_1^I}
=
|\theta-m_0|\frac{\tau_m}{(\tau_m+\tau_1^I)^2}\geq0,
\label{eq:app_update_margin_precision_derivative}
\end{equation}
with strict inequality when $\theta\neq m_0$. If $d\tau_1^I(a)/da>0$, then
\begin{equation}
\frac{dM}{da}
=
|\theta-m_0|\frac{\tau_m}{(\tau_m+\tau_1^I(a))^2}\frac{d\tau_1^I(a)}{da}\geq0,
\label{eq:app_update_margin_intensity_derivative}
\end{equation}
with strict inequality when $\theta\neq m_0$.

It remains to show that the update margin strictly decreases with $K$. From \eqref{eq:app_inaction_threshold}, let $z(K)=h^{-1}(h(1)-K)$. Since $h$ is strictly increasing, $z'(K)=-1/h'(z(K))<0$. Therefore,
\begin{equation}
\frac{d\Delta_K}{dK}
=
\frac{-z'(K)}{2\sqrt{1-z(K)}}>0.
\label{eq:app_threshold_derivative}
\end{equation}
Since $M=\omega^I(a)|\theta-m_0|-\Delta_K$, the update margin is strictly decreasing in $K$. \Halmos

\section*{Appendix B: Additional Tables and Robustness Checks}

\subsection*{Category-Level Distributions of Demographic Variables}

To supplement the summary balance tests presented in the main text , Table \ref{tab:appendix_demographics_distribution} provides a granular breakdown of the categorical demographic distributions across the treatment and control groups. We examine gender, community type, city level, and device price range to ensure that no specific demographic segment is disproportionately represented in either experimental condition. For confidentiality and data security purposes, the specific labels for sub-categories within community type, city level, and price range have been anonymized and randomly permuted. This detailed view confirms the integrity of the random assignment, as the proportions remain highly consistent across all 100,000 users in each group.

The results demonstrate that the absolute differences between the treatment and control shares are negligible for every category. For example, the maximum absolute difference in gender distribution is only 0.040\%, and the largest discrepancy across any demographic dimension is 0.341\%. These minimal variations across a wide array of user characteristics, including socio-economic indicators like device price range, provide strong evidence that the two groups are statistically identical prior to the intervention. Consequently, we can confidently attribute any subsequent divergence in late-night usage or overall engagement to the causal impact of the "sleep reminder" campaign rather than pre-existing differences in the user population.

\begin{table}[h!]
\centering
\scriptsize
\caption{Category-Level Distributions of Demographic Variables}
\label{tab:appendix_demographics_distribution}
\begin{threeparttable}

\begin{tabular}{lccc}
\toprule
\multicolumn{4}{c}{\textbf{Gender}} \\
\midrule
Category & Treatment (\%) & Control (\%) & Abs. Diff. (\%) \\
\midrule
Female & 46.894 & 46.934 & 0.040 \\
Male & 53.030 & 52.997 & 0.033 \\
Unknown & 0.076 & 0.069 & 0.007 \\
\bottomrule
\end{tabular}

\vspace{0.8em}

\begin{tabular}{lccc}
\toprule
\multicolumn{4}{c}{\textbf{Community Type}} \\
\midrule
Category & Treatment (\%) & Control (\%) & Abs. Diff. (\%) \\
\midrule
A & 0.069 & 0.056 & 0.013 \\
B & 28.281 & 28.380 & 0.099 \\
C & 19.391 & 19.461 & 0.070 \\
D & 38.394 & 38.231 & 0.163 \\
E & 13.865 & 13.872 & 0.007 \\
\bottomrule
\end{tabular}

\vspace{0.8em}

\begin{tabular}{lccc}
\toprule
\multicolumn{4}{c}{\textbf{City Level}} \\
\midrule
Category & Treatment (\%) & Control (\%) & Abs. Diff. (\%) \\
\midrule
A & 14.239 & 14.269 & 0.030 \\
B & 23.245 & 23.114 & 0.131 \\
C & 11.838 & 11.754 & 0.084 \\
D & 4.383 & 4.456 & 0.073 \\
E & 23.056 & 23.397 & 0.341 \\
F & 22.411 & 22.239 & 0.172 \\
G & 0.828 & 0.771 & 0.057 \\
\bottomrule
\end{tabular}

\vspace{0.8em}

\begin{tabular}{lccc}
\toprule
\multicolumn{4}{c}{\textbf{Device Price Range}} \\
\midrule
Category & Treatment (\%) & Control (\%) & Abs. Diff. (\%) \\
\midrule
A & 27.215 & 27.330 & 0.115 \\
B & 15.134 & 15.326 & 0.192 \\
C & 5.666 & 5.691 & 0.025 \\
D & 3.856 & 3.803 & 0.053 \\
E & 3.144 & 3.093 & 0.051 \\
F & 44.985 & 44.757 & 0.228 \\
\bottomrule
\end{tabular}

\begin{tablenotes}[flushleft]
\scriptsize
\item \textit{Notes:} This table reports category-level distributions of demographic variables used in the randomization check.
For confidentiality reasons, category labels are anonymized and randomly permuted within each variable.
\end{tablenotes}

\end{threeparttable}
\end{table}

\subsection*{Alternative Placebo Test Using Pre-Experiment Never-Nighters}

As a complement to the main placebo test in Table~\ref{tab:placebo_never_night}, which defines never-night users based on their behavior during the experimental period, Table~\ref{tab:placebo_never_night_preexp} reports an alternative specification in which never-nighters are identified using the pre-experiment week. Specifically, this subsample consists of users who had zero night usage during the one-week period prior to the experiment. Because this definition is based entirely on pre-treatment behavior, it avoids any concern about conditioning on a post-treatment outcome. The trade-off is that some of these users may have used the app during late-night hours during the experiment itself.

The results are broadly consistent with the main placebo test: neither App Usage nor Video Duration shows a statistically significant treatment effect for this subgroup. Interestingly, the coefficient on Night Usage is negative and marginally significant ($p < 0.1$), suggesting that among users who did not stay up late before the experiment, those assigned to the treatment group may have experienced a slight reduction in late-night usage during the campaign. While this point estimate should be interpreted cautiously given that it is only marginally significant and the subgroup definition does not guarantee the absence of treatment exposure, it provides suggestive evidence that, for at least some users, the sleep reminder content may have had a modest deterrent effect on late-night engagement, which is consistent with the campaign's original intent.

\begin{table}[h!]
\centering
\scriptsize
\caption{Alternative Placebo Test: Treatment Effects on Pre-Experiment Never-Night Users}
\label{tab:placebo_never_night_preexp}
\begin{threeparttable}
\begin{tabular}{lccc}
\toprule
& App Usage & Night Usage & Video Duration \\
\midrule
Treatment & 0.0082 & -0.0013$^{*}$ & -0.0028 \\
& (0.017) & (0.001) & (0.011) \\
Constant & 2.6626$^{***}$ & 0.0540$^{***}$ & 1.4765$^{***}$ \\
& (0.012) & (0.001) & (0.008) \\
\midrule
Observations & 533,448 & 533,448& 533,448 \\
\bottomrule
\end{tabular}
\begin{tablenotes}
\scriptsize \item \textit{Notes:} This table presents the results of OLS regressions analyzing the treatment effect on users who had zero night usage during the one-week pre-experimental period, i.e., users who were never active during late-night hours (1--5 AM) in the week before the experiment began. Unlike the main placebo test in Table~\ref{tab:placebo_never_night}, which conditions on during-experiment behavior, this definition is based entirely on pre-treatment behavior. The analysis is conducted at the user-day level during the experimental week. Because these users were not active at night before the experiment, the intervention is unlikely to have affected them. However, because some users in this subgroup may have begun using the app at night during the experiment, Night Usage is not mechanically zero and is therefore included as a dependent variable. Standard errors are in parentheses. $^{*}$ p $<$ 0.1, $^{**}$ p $<$ 0.05, $^{***}$ p $<$ 0.01.
\end{tablenotes}
\end{threeparttable}
\end{table}

\subsection*{Additional Placebo: Top-of-Feed Should Remain Unaffected}
\begin{figure}[t!]
    \centering
    \includegraphics[width=0.78\textwidth]{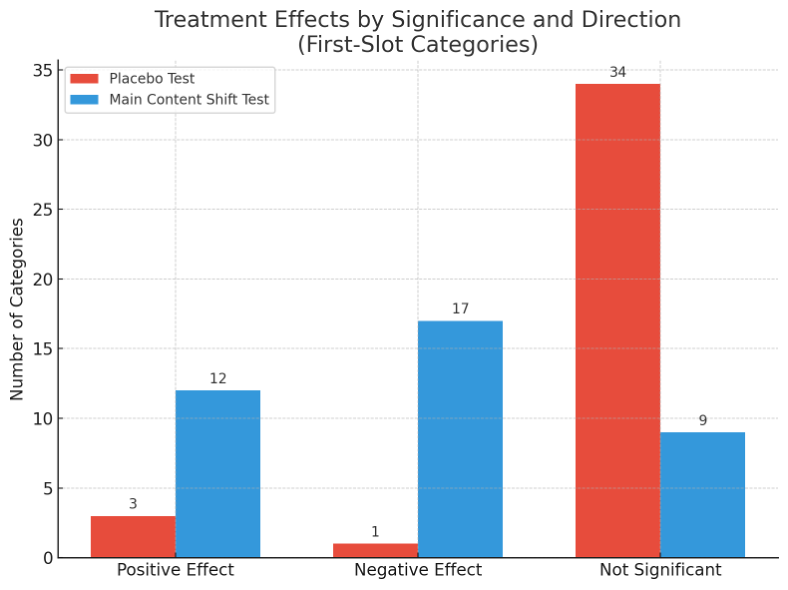}
    \caption{Placebo Test: Top-of-Feed Should Remain Unaffected}
    \label{fig:firstslot_placebo}
    \begin{minipage}{0.92\textwidth}
        \scriptsize
        \textit{Notes:}  The figure counts content categories by effect direction and significance at the very top of the feed. 
        Red bars are computed from day-1, session-level data using the video ranked first in each recommendation session for every user. 
        Blue bars report the main content-shift test based on daily aggregates of the top twenty recommendations during the experimental week. 
        More than 99 percent of reminder videos appear after the first position, so a valid manipulation should not move first-slot content on day one. 
        Interpretation should be cautious because the main test relies on daily top-twenty aggregates while boosting is applied at the session level. 
        A user can have multiple sessions in a day, so once a reminder video appears after the twentieth position in an early session, later sessions that day already reflect a treated environment. 
        Most of a user’s daily top-twenty slots may therefore occur after the treatment has taken effect. 
        Even under this conservative bias, the figure shows no systematic movement at the very top of the feed.

    \end{minipage}
\end{figure}

The second placebo test examines whether the intervention mechanically distorted the feed ranking process beyond the targeted positions. Over 99 percent of sleep-reminder videos appeared beyond the first slot in each recommendation session. If the treatment works only when a reminder video is actually seen, it should not affect the first-slot content, particularly on the first day of the experiment. Figure \ref{fig:firstslot_placebo} compares the number of content categories that show significant increases or decreases in exposure between the placebo test and the actual treatment. The results show no systematic difference across the two, suggesting that the top-of-feed content remained unaffected by the manipulation.  

This test, however, should be interpreted with caution. The available data record only the daily top-20 recommended videos per user, whereas the treatment was applied at the session level. A user could have multiple sessions in one day; if the reminder video appeared after the 20th slot in the first session, later sessions that day would already reflect the post-treatment environment. As a result, up to 80 to 90 percent of a user’s daily ``top 20” slots might already reflect exposure under treated conditions. Even with this conservative bias, the absence of measurable effects at the top of the feed reinforces that the manipulation was localized to the intended recommendation rule.

\subsection*{Long-Term Effects on Engagement}
\begin{table}[t!]
\centering
\scriptsize
\caption{Long-Term Treatment Effects Over Time: App Usage}
\label{tab:longterm_appusage}
\begin{threeparttable}

\begin{tabular}{lccccccccc}
\toprule
\multicolumn{10}{l}{\textbf{Panel A: OLS (ITT, DV $\sim$ Treatment)}} \\
\midrule
 & \textbf{Week 0} & \textbf{Week 1} & \textbf{Week 2} & \textbf{Week 3} & \textbf{Week 4} & \textbf{Week 5} & \textbf{Week 6} & \textbf{Week 7} & \textbf{Week 8} \\
\cmidrule(lr){2-10}
Treatment & 0.0246$^{***}$ & 0.0085$^{*}$ & 0.0075$^{*}$ & 0.0089$^{**}$ & 0.0087$^{**}$ & 0.0086$^{**}$ & 0.0077$^{*}$ & 0.0058 & 0.0035 \\
          & (0.0052) & (0.0045) & (0.0044) & (0.0043) & (0.0043) & (0.0043) & (0.0043) & (0.0043) & (0.0044) \\
Constant  & 1.1872$^{***}$ & 0.9932$^{***}$ & 0.9395$^{***}$ & 0.9246$^{***}$ & 0.9197$^{***}$ & 0.9121$^{***}$ & 0.9035$^{***}$ & 0.8985$^{***}$ & 0.9187$^{***}$ \\
          & (0.0036) & (0.0032) & (0.0031) & (0.0031) & (0.0030) & (0.0031) & (0.0031) & (0.0030) & (0.0031) \\
\midrule
Relative Effect (\%) & 2.07 & 0.86 & 0.80 & 0.97 & 0.94 & 0.94 & 0.85 & -- & -- \\
Observations & 200{,}000 & 200{,}000 & 200{,}000 & 200{,}000 & 200{,}000 & 200{,}000 & 200{,}000 & 200{,}000 & 200{,}000 \\
\bottomrule
\end{tabular}

\vspace{0.8em}

\begin{tabular}{lccccccccc}
\toprule
\multicolumn{10}{l}{\textbf{Panel B: IV (LATE, DV $\sim$ Treated)}} \\
\midrule
 & \textbf{Week 0} & \textbf{Week 1} & \textbf{Week 2} & \textbf{Week 3} & \textbf{Week 4} & \textbf{Week 5} & \textbf{Week 6} & \textbf{Week 7} & \textbf{Week 8} \\
\cmidrule(lr){2-10}
Treated   & 0.1517$^{***}$ & 0.0525$^{*}$ & 0.0462$^{*}$ & 0.0551$^{**}$ & 0.0533$^{**}$ & 0.0531$^{**}$ & 0.0476$^{*}$ & 0.0356 & 0.0213 \\
          & (0.0315) & (0.0276) & (0.0269) & (0.0267) & (0.0264) & (0.0266) & (0.0266) & (0.0264) & (0.0271) \\
Constant  & 1.1838$^{***}$ & 0.9920$^{***}$ & 0.9385$^{***}$ & 0.9234$^{***}$ & 0.9185$^{***}$ & 0.9109$^{***}$ & 0.9025$^{***}$ & 0.8977$^{***}$ & 0.9182$^{***}$ \\
          & (0.0041) & (0.0036) & (0.0035) & (0.0035) & (0.0035) & (0.0035) & (0.0035) & (0.0035) & (0.0036) \\
\midrule
Relative Effect (\%) & 12.82 & 5.29 & 4.92 & 5.97 & 5.81 & 5.83 & 5.27 & -- & -- \\
Observations & 200{,}000 & 200{,}000 & 200{,}000 & 200{,}000 & 200{,}000 & 200{,}000 & 200{,}000 & 200{,}000 & 200{,}000 \\
\bottomrule
\end{tabular}

\begin{tablenotes}
\scriptsize
\item \textit{Notes:} This table reports treatment effects on user engagement over time following the experimental week. The analysis is at the user-week level. The dependent variable is standardized to have unit standard deviation. Each column represents a separate week (Week 0 is the experimental week). Panel A reports the intent-to-treat (ITT) effect from an OLS regression of the outcome on the treatment assignment dummy. Panel B reports the local average treatment effect (LATE) from a 2SLS regression, where random assignment to the treatment group is used as an instrument for whether the user was actually exposed to a reminder video. Treatment assignment and actual treatment receipt remain constant throughout the study period (first stage results are shown in the short-term analysis table). Robust standard errors in parentheses. $^{***}p<0.01$, $^{**}p<0.05$, $^{*}p<0.1$. Relative effects are computed as 100$\times$Coefficient / Constant and are omitted for non-significant coefficients (p $\geq$ 0.1).
\end{tablenotes}
\end{threeparttable}
\end{table}

\begin{table}[t!]
\centering
\scriptsize
\caption{Long-Term Treatment Effects Over Time: Night Usage}
\label{tab:longterm_nightusage}
\begin{threeparttable}

\begin{tabular}{lccccccccc}
\toprule
\multicolumn{10}{l}{\textbf{Panel A: OLS (ITT, DV $\sim$ Treatment)}} \\
\midrule
 & \textbf{Week 0} & \textbf{Week 1} & \textbf{Week 2} & \textbf{Week 3} & \textbf{Week 4} & \textbf{Week 5} & \textbf{Week 6} & \textbf{Week 7} & \textbf{Week 8} \\
\cmidrule(lr){2-10}
Treatment & 0.0725$^{***}$ & 0.0350$^{***}$ & 0.0301$^{***}$ & 0.0291$^{***}$ & 0.0241$^{***}$ & 0.0208$^{***}$ & 0.0225$^{***}$ & 0.0190$^{***}$ & 0.0166$^{***}$ \\
          & (0.0048) & (0.0047) & (0.0046) & (0.0045) & (0.0044) & (0.0044) & (0.0044) & (0.0042) & (0.0043) \\
Constant  & 0.5245$^{***}$ & 0.5046$^{***}$ & 0.4900$^{***}$ & 0.4837$^{***}$ & 0.4876$^{***}$ & 0.4717$^{***}$ & 0.4700$^{***}$ & 0.4594$^{***}$ & 0.4599$^{***}$ \\
          & (0.0032) & (0.0033) & (0.0032) & (0.0031) & (0.0031) & (0.0031) & (0.0030) & (0.0030) & (0.0030) \\
\midrule
Relative Effect (\%) & 13.82 & 6.93 & 6.14 & 6.02 & 4.95 & 4.41 & 4.80 & 4.15 & 3.60 \\
Observations & 200{,}000 & 200{,}000 & 200{,}000 & 200{,}000 & 200{,}000 & 200{,}000 & 200{,}000 & 200{,}000 & 200{,}000 \\
\bottomrule
\end{tabular}

\vspace{0.8em}

\begin{tabular}{lccccccccc}
\toprule
\multicolumn{10}{l}{\textbf{Panel B: IV (LATE, DV $\sim$ Treated)}} \\
\midrule
 & \textbf{Week 0} & \textbf{Week 1} & \textbf{Week 2} & \textbf{Week 3} & \textbf{Week 4} & \textbf{Week 5} & \textbf{Week 6} & \textbf{Week 7} & \textbf{Week 8} \\
\cmidrule(lr){2-10}
Treated   & 0.4469$^{***}$ & 0.2156$^{***}$ & 0.1856$^{***}$ & 0.1795$^{***}$ & 0.1488$^{***}$ & 0.1282$^{***}$ & 0.1390$^{***}$ & 0.1174$^{***}$ & 0.1022$^{***}$ \\
          & (0.0282) & (0.0286) & (0.0278) & (0.0273) & (0.0269) & (0.0268) & (0.0267) & (0.0260) & (0.0260) \\
Constant  & 0.5144$^{***}$ & 0.4997$^{***}$ & 0.4858$^{***}$ & 0.4796$^{***}$ & 0.4842$^{***}$ & 0.4688$^{***}$ & 0.4669$^{***}$ & 0.4568$^{***}$ & 0.4576$^{***}$ \\
          & (0.0037) & (0.0038) & (0.0037) & (0.0036) & (0.0035) & (0.0035) & (0.0035) & (0.0034) & (0.0034) \\
\midrule
Relative Effect (\%) & 86.88 & 43.14 & 38.20 & 37.42 & 30.73 & 27.35 & 29.77 & 25.71 & 22.33 \\
Observations & 200{,}000 & 200{,}000 & 200{,}000 & 200{,}000 & 200{,}000 & 200{,}000 & 200{,}000 & 200{,}000 & 200{,}000 \\
\bottomrule
\end{tabular}

\begin{tablenotes}
\scriptsize
\item \textit{Notes:} This table reports treatment effects on user engagement over time following the experimental week. The analysis is at the user-week level. The dependent variable is standardized to have unit standard deviation. Each column represents a separate week (Week 0 is the experimental week). Panel A reports the intent-to-treat (ITT) effect from an OLS regression of the outcome on the treatment assignment dummy. Panel B reports the local average treatment effect (LATE) from a 2SLS regression, where random assignment to the treatment group is used as an instrument for whether the user was actually exposed to a reminder video. Treatment assignment and actual treatment receipt remain constant throughout the study period (first stage results are shown in the short-term analysis table). Robust standard errors in parentheses. $^{***}p<0.01$, $^{**}p<0.05$, $^{*}p<0.1$. Relative effects are computed as 100$\times$Coefficient / Constant and are omitted for non-significant coefficients (p $\geq$ 0.1).
\end{tablenotes}
\end{threeparttable}
\end{table}

\begin{table}[t!]
\centering
\scriptsize
\caption{Long-Term Treatment Effects Over Time: Video Duration}
\label{tab:longterm_photoduration}
\begin{threeparttable}

\begin{tabular}{lccccccccc}
\toprule
\multicolumn{10}{l}{\textbf{Panel A: OLS (ITT, DV $\sim$ Treatment)}} \\
\midrule
 & \textbf{Week 0} & \textbf{Week 1} & \textbf{Week 2} & \textbf{Week 3} & \textbf{Week 4} & \textbf{Week 5} & \textbf{Week 6} & \textbf{Week 7} & \textbf{Week 8} \\
\cmidrule(lr){2-10}
Treatment & 0.0302$^{***}$ & 0.0117$^{***}$ & 0.0105$^{**}$ & 0.0107$^{**}$ & 0.0109$^{**}$ & 0.0115$^{***}$ & 0.0101$^{**}$ & 0.0081$^{*}$ & 0.0059 \\
          & (0.0053) & (0.0045) & (0.0043) & (0.0043) & (0.0043) & (0.0043) & (0.0043) & (0.0043) & (0.0044) \\
Constant  & 1.1973$^{***}$ & 0.9874$^{***}$ & 0.9221$^{***}$ & 0.9129$^{***}$ & 0.9144$^{***}$ & 0.9034$^{***}$ & 0.8951$^{***}$ & 0.8928$^{***}$ & 0.9140$^{***}$ \\
          & (0.0037) & (0.0032) & (0.0031) & (0.0030) & (0.0030) & (0.0030) & (0.0030) & (0.0030) & (0.0031) \\
\midrule
Relative Effect (\%) & 2.52 & 1.19 & 1.13 & 1.18 & 1.19 & 1.27 & 1.13 & 0.91 & -- \\
Observations & 200{,}000 & 200{,}000 & 200{,}000 & 200{,}000 & 200{,}000 & 200{,}000 & 200{,}000 & 200{,}000 & 200{,}000 \\
\bottomrule
\end{tabular}

\vspace{0.8em}

\begin{tabular}{lccccccccc}
\toprule
\multicolumn{10}{l}{\textbf{Panel B: IV (LATE, DV $\sim$ Treated)}} \\
\midrule
 & \textbf{Week 0} & \textbf{Week 1} & \textbf{Week 2} & \textbf{Week 3} & \textbf{Week 4} & \textbf{Week 5} & \textbf{Week 6} & \textbf{Week 7} & \textbf{Week 8} \\
\cmidrule(lr){2-10}
Treated   & 0.1859$^{***}$ & 0.0722$^{***}$ & 0.0645$^{**}$ & 0.0663$^{**}$ & 0.0671$^{**}$ & 0.0708$^{***}$ & 0.0625$^{**}$ & 0.0499$^{*}$ & 0.0361 \\
          & (0.0321) & (0.0276) & (0.0266) & (0.0264) & (0.0263) & (0.0263) & (0.0264) & (0.0263) & (0.0272) \\
Constant  & 1.1931$^{***}$ & 0.9858$^{***}$ & 0.9206$^{***}$ & 0.9114$^{***}$ & 0.9129$^{***}$ & 0.9018$^{***}$ & 0.8936$^{***}$ & 0.8917$^{***}$ & 0.9132$^{***}$ \\
          & (0.0042) & (0.0036) & (0.0035) & (0.0035) & (0.0035) & (0.0035) & (0.0035) & (0.0035) & (0.0036) \\
\midrule
Relative Effect (\%) & 15.58 & 7.33 & 7.00 & 7.27 & 7.35 & 7.85 & 6.99 & 5.60 & -- \\
Observations & 200{,}000 & 200{,}000 & 200{,}000 & 200{,}000 & 200{,}000 & 200{,}000 & 200{,}000 & 200{,}000 & 200{,}000 \\
\bottomrule
\end{tabular}

\begin{tablenotes}
\scriptsize
\item \textit{Notes:} This table reports treatment effects on user engagement over time following the experimental week. The analysis is at the user-week level. The dependent variable is standardized to have unit standard deviation. Each column represents a separate week (Week 0 is the experimental week). Panel A reports the intent-to-treat (ITT) effect from an OLS regression of the outcome on the treatment assignment dummy. Panel B reports the local average treatment effect (LATE) from a 2SLS regression, where random assignment to the treatment group is used as an instrument for whether the user was actually exposed to a reminder video. Treatment assignment and actual treatment receipt remain constant throughout the study period (first stage results are shown in the short-term analysis table). Robust standard errors in parentheses. $^{***}p<0.01$, $^{**}p<0.05$, $^{*}p<0.1$. Relative effects are computed as 100$\times$Coefficient / Constant and are omitted for non-significant coefficients (p $\geq$ 0.1).
\end{tablenotes}
\end{threeparttable}
\end{table}

\begin{table}[t!]
\centering
\scriptsize
\caption{Long-Term Treatment Effects Over Time: Live Duration}
\label{tab:longterm_liveduration}
\begin{threeparttable}

\begin{tabular}{lccccccccc}
\toprule
\multicolumn{10}{l}{\textbf{Panel A: OLS (ITT, DV $\sim$ Treatment)}} \\
\midrule
 & \textbf{Week 0} & \textbf{Week 1} & \textbf{Week 2} & \textbf{Week 3} & \textbf{Week 4} & \textbf{Week 5} & \textbf{Week 6} & \textbf{Week 7} & \textbf{Week 8} \\
\cmidrule(lr){2-10}
Treatment & 0.0020 & -0.0006 & 0.0018 & 0.0018 & 0.0024 & 0.0017 & 0.0031 & 0.0026 & 0.0009 \\
          & (0.0051) & (0.0046) & (0.0045) & (0.0044) & (0.0042) & (0.0044) & (0.0043) & (0.0043) & (0.0043) \\
Constant  & 0.3926$^{***}$ & 0.3531$^{***}$ & 0.3471$^{***}$ & 0.3349$^{***}$ & 0.3176$^{***}$ & 0.3258$^{***}$ & 0.3236$^{***}$ & 0.3193$^{***}$ & 0.3260$^{***}$ \\
          & (0.0036) & (0.0032) & (0.0032) & (0.0031) & (0.0030) & (0.0031) & (0.0030) & (0.0030) & (0.0031) \\
\midrule
Relative Effect (\%) & -- & -- & -- & -- & -- & -- & -- & -- & -- \\
Observations & 200{,}000 & 200{,}000 & 200{,}000 & 200{,}000 & 200{,}000 & 200{,}000 & 200{,}000 & 200{,}000 & 200{,}000 \\
\bottomrule
\end{tabular}

\vspace{0.8em}

\begin{tabular}{lccccccccc}
\toprule
\multicolumn{10}{l}{\textbf{Panel B: IV (LATE, DV $\sim$ Treated)}} \\
\midrule
 & \textbf{Week 0} & \textbf{Week 1} & \textbf{Week 2} & \textbf{Week 3} & \textbf{Week 4} & \textbf{Week 5} & \textbf{Week 6} & \textbf{Week 7} & \textbf{Week 8} \\
\cmidrule(lr){2-10}
Treated   & 0.0121 & -0.0039 & 0.0110 & 0.0111 & 0.0147 & 0.0105 & 0.0191 & 0.0159 & 0.0053 \\
          & (0.0312) & (0.0283) & (0.0279) & (0.0273) & (0.0262) & (0.0270) & (0.0268) & (0.0263) & (0.0267) \\
Constant  & 0.3923$^{***}$ & 0.3532$^{***}$ & 0.3469$^{***}$ & 0.3347$^{***}$ & 0.3172$^{***}$ & 0.3256$^{***}$ & 0.3231$^{***}$ & 0.3190$^{***}$ & 0.3259$^{***}$ \\
          & (0.0041) & (0.0037) & (0.0037) & (0.0036) & (0.0034) & (0.0036) & (0.0035) & (0.0035) & (0.0035) \\
\midrule
Relative Effect (\%) & -- & -- & -- & -- & -- & -- & -- & -- & -- \\
Observations & 200{,}000 & 200{,}000 & 200{,}000 & 200{,}000 & 200{,}000 & 200{,}000 & 200{,}000 & 200{,}000 & 200{,}000 \\
\bottomrule
\end{tabular}

\begin{tablenotes}
\scriptsize
\item \textit{Notes:} This table reports treatment effects on user engagement over time following the experimental week. The analysis is at the user-week level. The dependent variable is standardized to have unit standard deviation. Each column represents a separate week (Week 0 is the experimental week). Panel A reports the intent-to-treat (ITT) effect from an OLS regression of the outcome on the treatment assignment dummy. Panel B reports the local average treatment effect (LATE) from a 2SLS regression, where random assignment to the treatment group is used as an instrument for whether the user was actually exposed to a reminder video. Treatment assignment and actual treatment receipt remain constant throughout the study period (first stage results are shown in the short-term analysis table). Robust standard errors in parentheses. $^{***}p<0.01$, $^{**}p<0.05$, $^{*}p<0.1$. Relative effects are computed as 100$\times$Coefficient / Constant and are omitted for non-significant coefficients (p $\geq$ 0.1).
\end{tablenotes}
\end{threeparttable}
\end{table}

\begin{table}[t!]
    \centering
    \scriptsize
    \caption{Relative Effect Sizes of Treatment on Conversion Rates (\%)}
    \label{tab:category_conversion}
    \begin{threeparttable}
    \begin{tabular}{l | c c c}
        \toprule
        \textbf{Category} 
        & \textbf{Baseline (\%)} 
        & \multicolumn{2}{c}{\textbf{Relative Effect (\%)}} \\
        \cmidrule(lr){3-4}
        & 
        & \textbf{Week 0} 
        & \textbf{Week 1--8} \\
        \midrule
        Astrology \& Fortune-telling & 8.17 & 2.78*** & 1.71*** \\
        Oddities \& Wonders & 23.00 & 2.18*** & 1.33*** \\
        Fitness & 11.30 & 2.03*** & 0.98*** \\
        Humanities & 19.40 & 1.59*** & 1.06*** \\
        Anime / ACG & 21.82 & 1.50*** & 0.94*** \\
        Pets & 23.20 & 1.45*** & 0.96*** \\
        Science \& Law & 22.87 & 1.42*** & 1.21*** \\
        Military & 20.68 & 1.41*** & 1.00*** \\
        Sports & 26.69 & 1.37*** & 1.13*** \\
        Games & 20.38 & 1.26*** & 1.35*** \\
        Art & 27.05 & 1.25*** & 0.88*** \\
        Appearance / Looks & 14.88 & 1.24*** & 1.02*** \\
        Travel & 21.14 & 1.19*** & 0.86*** \\
        Celebrity Entertainment & 27.18 & 1.18*** & 0.84*** \\
        Health & 24.55 & 1.09*** & 0.52*** \\
        Beauty / Makeup & 21.45 & 1.05*** & 0.52*** \\
        Music & 26.95 & 1.04*** & 0.26*** \\
        Reading / Books & 24.10 & 0.80*** & 0.57*** \\
        Short Dramas & 35.85 & 0.77*** & 0.57*** \\
        Photography & 16.51 & 0.73** & 0.77*** \\
        Finance \& Business & 17.48 & 0.67** & 0.61*** \\
        Lifestyle & 31.11 & 0.66*** & 0.59*** \\
        Automobiles & 25.44 & 0.65*** & 0.33*** \\
        Film \& TV Shows & 39.67 & 0.64*** & 0.51*** \\
        Parenting \& Family & 29.44 & 0.61*** & 0.40*** \\
        Food & 31.57 & 0.53*** & 0.49*** \\
        Education & 18.03 & 0.50 & 0.32*** \\
        Current Affairs \& News & 14.07 & 0.45 & 0.11 \\
        Comedy & 41.71 & 0.44*** & 0.57*** \\
        Real Estate \& Home & 31.75 & 0.44** & 0.14* \\
        Rural / Agriculture & 28.16 & 0.43** & 0.32*** \\
        Civic News & 40.77 & 0.39*** & 0.54*** \\
        Selfies & 24.37 & 0.36* & 0.14 \\
        Dance & 23.19 & 0.33 & 0.13 \\
        Emotions / Relationships & 32.04 & 0.32* & 0.16* \\
        Fashion & 27.31 & 0.01 & -0.18** \\
        Others & 23.71 & -0.45* & -0.16* \\
        Tech \& Digital Products & 21.85 & -0.67** & -0.63*** \\
        \bottomrule
    \end{tabular}
    \begin{tablenotes}
    \scriptsize
    \item Note: This table shows the treatment's effects on the conversion rate (played / recommended) for different content categories. The analysis is at the user-day-category level. Baseline (\%)" shows the average conversion rate for the control group. Relative Effect (\%)" columns report the ITT effect as a percentage, calculated from our main OLS specification as (coefficient on Treatment / baseline) $\times$ 100. The effects are shown separately for the experimental week (Week 0) and the following eight weeks (Week 1-8). Positive values indicate higher conversion, while negative values indicate lower conversion.$^{*}$p$<$0.1; $^{**}$p$<$0.05; $^{***}$p$<$0.01.
    \end{tablenotes}
    \end{threeparttable}
\end{table}

\end{document}